\documentclass[10pt,emulateapj,apj]{emulateapj}
\usepackage{enumerate}
\usepackage{amsmath}

\shorttitle{Isolated Galaxy Models}
\shortauthors{Hwang et al.}

\begin{document}
\title{THE INITIAL CONDITIONS AND EVOLUTION OF ISOLATED GALAXY MODELS: EFFECTS OF THE HOT GAS HALO}
\author{Jeong-Sun Hwang$^1$, Changbom Park$^1$, \& Jun-Hwan Choi$^{2}$}
\affil{$^1$ School of Physics, Korea Institute for Advanced Study, Heogiro 85, Seoul 130-722, Korea \\
$^2$Department of Physics and Astronomy, University of Kentucky, Lexington, KY, 40506-0055, U.S.A.; jhchoi@pa.uky.edu\\}
\email{jhchoi@pa.uky.edu}


\begin{abstract}

We construct several Milky Way-like galaxy models
containing a gas halo
(as well as gaseous and stellar disks, a dark matter halo, and a stellar bulge)
following either an isothermal or an NFW density profile
with varying mass and initial spin.
In addition, galactic winds associated with star formation are tested
in some of the simulations.
We evolve these isolated galaxy models using
the GADGET-3 $N$-body/hydrodynamic simulation code, paying particular attention
to the effects of the gaseous halo on the evolution.
We find that the evolution of the models is strongly affected
by the adopted gas halo component,
particularly in the gas dissipation
and the star formation activity in the disk.
The model without a gas halo shows an increasing star formation rate (SFR)
at the beginning of the simulation
for some hundreds of millions of years
and then a continuously decreasing rate to the end of the run at 3~Gyr.
Whereas the SFRs in the models with a gas halo,
depending on the density profile and the total mass of the gas halo,
emerge to be either relatively flat throughout the simulations
or increasing until the middle of the run
over a gigayear, then decreasing to the end.
The models with the more centrally concentrated NFW gas halo
show overall higher SFRs
than those with the isothermal gas halo of the equal mass.
The gas accretion from the halo onto the disk
also occurs more in the models with the NFW gas halo,
however, this is shown to take place mostly in the inner part of the disk
and not to contribute significantly to the star formation
unless the gas halo has very high density at the central part.
The rotation of a gas halo is found to make
SFR lower in the model.
The SFRs in the runs including galactic winds
are found to be lower than the same runs but without winds.
We conclude that the effects of a hot gaseous halo on the evolution
of galaxies are generally too significant to be simply ignored.
We also expect that more hydrodynamical processes in galaxies
could be understood through numerical simulations
employing both gas disk and gas halo components.

\end{abstract}




\section{INTRODUCTION}

Under the hierarchical galaxy formation scenario,
galaxy interactions play a key role in the evolution of galaxies (e.g. \citealt{Toomre1972,Barnes1992}).
Owing to the nonlinear nature of the galaxy interactions,
theoretical studies mostly rely on $N$-body/hydrodynamic simulations.
The initial conditions of the simulations often adopt isolated galaxy models.
Galaxy models in many galaxy interaction simulations include dark matter halo,
gas disk, stellar disk, and stellar bulge \citep{Hernquist1993,Springel2005}.
Unfortunately, generating an isolated stable galaxy model is not trivial procedure.
It is required to solve complicated orbit integrations to make stable stellar systems \citep{Osipkov1979,Merritt1985,Kazabtzidis2004}.
The interactions among the different components significantly modify the very initial setups \citep{Blumenthal1984}.
Furthermore, it is needed to satisfy the initially hydrostatic condition of the gas components.
In order to isolate the galaxy interaction effects,
it is necessary to carefully build the initial galaxy model
and understand the evolution of the initial conditions, if there are any.
However, the basic steps to construct various types of initial galaxy models
are often omitted or discussed shortly.
The detailed comparison studies of the evolution of different galaxy models in isolation
are also not found sufficiently in the literature.

Moreover, it has been reported that some spiral galaxies like our own Milky Way as well as elliptical galaxies possess hot diffuse gas in the halo.
Importance of hot halo gas in the evolution of galaxies has been also evidenced by
\citet{Park2008}, \citet{Park_Choi2009}, \citet{Park_Hwang2009}, and \citet{Hwang2009}
who studied the effects of galaxy interactions on galaxy properties such as morphology, color, and star formation activity among others at various environment and redshifts.
It has been consistently suggested by these studies that the hydrodynamic interaction between galaxies through exchange of cold gas and/or impact of hot halo gas makes a crucial role in the evolution of galaxy color and star formation activity.
However, only a few numerical studies taking account of halo gas in the models have been presented.
Recently, \citet{Moster2011}
performed hydrodynamic simulations of major mergers of disk galaxies,
including for the first time a gradually cooling hot gas halo
in their galaxy models.
They studied the impact of a gaseous halo and galactic winds
on the star formation activity during mergers
and found that the gas halo component strongly affects
the kinematics and internal structure of the merger remnants
as well as the star formation history on the course of the merger.
In a follow-up study \citep{Moster2012},
they also investigated the role of a hot gas component
on disk thickening during minor merger and
showed the dependency of the final scale height
on the mass of the hot halo, the efficiency of the winds,
and the merger mass ratio.
They found that the cooling of the hot gaseous halo regenerates the thin disk,
so that the remnants of 1:10 mergers exhibit a newly formed thin disk.
These findings suggest that the existence of a hot gaseous halo significantly affects
the results of galaxy interactions.
It may be necessary to reinterpret many previous
galactic scale simulation results
that did not consider a hot gaseous halo.

In this paper, we present detailed descriptions of
the various choices of the isolated Milky Way-like galaxy models.
We also study the effects of
the gas halo component with different properties,
such as the cumulative mass profile, total mass,
and rotation of the gas halo on their evolution.
This paper is the first in a series of the galaxy interaction study
that includes a hot gas halo in the initial galaxy model with a hope to explain the observational studies by
\citet{Park2008}, \citet{Park_Choi2009}, \citet{Park_Hwang2009}, \citet{Hwang2009}, and \citet{Deng2012}.

Recent galaxy formation studies suggest that the galactic outflow is an essential
mechanism for balance between gas accretion and galaxy growth (e.g. \citealt{Dekel1986,Springel2003,Choi2011,Hopkins2011}).
It also regulates the star formation rate of galaxies.
In our study, we also include galactic outflow driven by supernovae
with different wind parameters
to study the effects of a gas halo and galactic winds
on the evolution of isolated galaxies.

In Sec.~2, we give a brief account of the ZENO software package
and explain our galaxy models and their initial parameters.
We describe in Sec.~3 the simulation code GADGET-3
and the results of evolution of the models.
Finally, we summarize and discuss our main results in Sec.~4.
The step-by-step procedure of constructing our galaxy models
is presented in Appendix~A,
and the long-term evolution of one of our models in Appendix~B.


\section{GALAXY MODELS}


\subsection{Numerical Code}

To generate the initial galaxy models,
we use the ZENO software package (version 008).
ZENO is a collection of programs for $N$-body/SPH simulations and analysis,
provided by J. E. Barnes\footnote{http://www.ifa.hawaii.edu/$\sim$barnes/software.html}.

ZENO allows one to build multiple spherical and disk components
in mutual equilibrium
with user-specified density profiles in collisionless or gaseous form,
that fulfills our needs for constructing various galaxy models.
We use about a dozen programs in the ZENO package
to complete our multi-component galaxy models.
The procedure of constructing the models using those programs
is described in Appendix~A in detail.

The ZENO programs use dimensionless units with Newton's constant $G$ = 1.
We set the total disk mass (including both stars and gas) as the unit mass
and the radial scale length of the stellar disk
as the unit length in all models.
We set the components of our models
to follow those of the Milky Way Galaxy, and
the units of length, mass, and time are
3.5~kpc, $5 \times 10^{10}$ M$_{\odot}$, and 14~Myr, respectively.
The unit velocity then corresponds to 247~km~s$^{-1}$.


\subsection{Density Profiles}

In our models both star and gas disks
have an exponential surface density profile and a sech$^2$ vertical profile
(c.f. \citealt{Barnes2009}):
\begin{equation}
  \rho_{\rm dc}(R,z) =
    \frac{M_{\rm dc}}{4 \pi a_{\rm dc}^2 z_{\rm dc}} \,
      e^{- R / a_{\rm dc}} \,
        \mathrm{sech}^2 \left( \frac{\it z}{\it z_{\rm dc}} \right) \, ,
  \label{eq1}
\end{equation}
where `$\rm c$' stands for either `$\rm s$' for the star disk component
or `$\rm g$' for the gas disk component.
$R$ is the cylindrical radius $R = (x^2+y^2)^{1/2}$ and
$M_{\rm dc}$ is the mass of the star or the gas disk.
$a_{\rm dc}$ is the radial scale length and
$z_{\rm dc}$ is the vertical scale height of the disk.
A few distant particles farther than a cut-off radius ($b_{\rm dc}$)
are removed.
The gas disk is radially more extended than the star disk in our models.
The radial and vertical scales of the gas disk are chosen as
$a_{\rm dg} = 2.5 \times a_{\rm ds}$ and $z_{\rm dg} = z_{\rm ds}$.

The mass density of the DM halo of all models
and the gas halo of some models
follows an NFW profile (\citealt{Navarro1996}).
Since the cumulative mass of the profile diverges at large radii,
an exponential taper is applied at radii larger than $b_{\rm hc}$:
\begin{equation}
\rho_{\rm hc}(r) =
  \left\{
  \begin{array}{ll}
   \displaystyle
   \frac{M_{\rm hc}( a_{\rm hc})}{4 \pi (\ln(2) - \frac{1}{2})} \,
   \frac{1}{r (r + a_{\rm hc})^2} &
	    {\rm for} \,\, r \le b_{\rm hc} \, , \\ [0.4cm]
   \displaystyle
   \rho_{\rm hc}^{*} \, \left(\frac{b_{\rm hc}}{r}\right)^2 \,
	    e^{-2 \beta (r / b_{\rm hc} -1)} &
	    {\rm for} \,\, r > b_{\rm hc} \, , \\
  \end{array}
  \right.
\label{eq2}
\end{equation}
where `$\rm c$' represents either `$\rm d$' for the DM halo component
or `$\rm g$' for the gas halo component.
$a_{\rm hc}$ is the radial scale of the DM or gas halo model and
$M_{\rm hc}(a_{\rm hc})$ is the mass of the halo within $a_{\rm hc}$.
The radial scales of both DM and gas halos
are chosen to be the same.
$\rho_{\rm hc}^{*}$ and $\beta$ are calculated by requiring
both $\rho_{\rm hc}(r)$ and $d \rho_{\rm hc} / d r$ to be
continuous at $b_{\rm hc}$.

The gas halo of the rest of our models
follows a non-singular isothermal profiles with truncation:
\begin{equation}
  \rho_{\rm hg}(r) =
	\displaystyle
          \frac{f_{\rm norm}  M_{\rm hg}}{2 {\pi} \sqrt{\pi} \, b_{\rm hg}} \,
          \frac{1}{r^2 + a_{\rm hg}^2} \,  e^{- (r / b_{\rm hg})^2}
	       \, ,  \\
  \label{eq3}
\end{equation}
where $a_{\rm hg}$ and $b_{\rm hg}$ are the radius of core and taper, respectively.
$f_{\rm norm}$ is a function of $a_{\rm hg}$ and $b_{\rm hg}$,
and $M_{\rm hg}$ is the total mass of the halo gas.

The mass density of the bulge, which consists of stars only,
in all models follows
a \citet{Hernquist1990} profile, with a cut-off at large radii
for efficient computation:
\begin{equation}
  \rho_{\rm b}(r) =
    \left\{
      \begin{array}{ll}
	\displaystyle
          \frac{a_{\rm b} M_{\rm b}}{2 \pi} \,
	      \frac{1}{r (a_{\rm b} + r)^{3}} \, &
	     {\rm for} \,\, r \le b_{\rm b} \, , \\ [0.4cm]
	\displaystyle
	  \rho_{\rm b}^{*} \, \left(\frac{b_{\rm b}}{r}\right)^2 \,
	      e^{-2 r / b_{\rm b}} \,  &
	     {\rm for} \,\, r > b_{\rm b} \, , \\
      \end{array}
    \right.
  \label{eq:hernquist}
\end{equation}
where $a_{\rm b}$ is the radial scale of the bulge model and
$b_{\rm b}$ is the radius at which truncation starts.
$M_{\rm b}$ is the mass of the bulge.
$\rho_{\rm b}^{*}$ is determined
by the continuity of $\rho_{\rm b}(r)$ at $r = b_{\rm b}$.
For $b_{\rm b} \gg a_{\rm b}$,
$d \rho_{\rm b} / d r$ is continuous at $r = b_{\rm b}$ as well.


\subsection{Initial Galaxy Models}

We construct seven different galaxy models in total.
Their basic properties are summarized in Tables~1 and 2.
We first build model~DHi as our fiducial model.
It contains a gas halo component following an isothermal density profile.
We then generate six more comparison models
without or with a gas halo of varying properties.

\begin{deluxetable*}{lcccl}
\tablecolumns{5}
\tablewidth{0pc}
\tablecaption{Initial galaxy models \label{tab01}}
\tablehead{
\colhead{Model name} &
\colhead{Gas halo model} &
\colhead{Gas halo rotation} &
\colhead{$f_{\rm hg}$\tablenotemark{a}}&
\colhead{$f_{\rm dg}$\tablenotemark{b}}
}
\startdata
DHi    & isothermal & $\cdots$                     & 0.01      & 0.12 \\
DHi-f5 & isothermal & $\cdots$                     & 0.05      & 0.12 \\
DHir   & isothermal & gas disk rotation$\times$0.5 & 0.01      & 0.12 \\
\cline{1-5}\\
DHn    & NFW        & $\cdots$                     & 0.01      & 0.12 \\
DHn-f5 & NFW        & $\cdots$                     & 0.05      & 0.12  \\
\cline{1-5}\\
D     & $\cdots$    & $\cdots$                     & $\cdots$  & 0.12 \\
\cline{1-5}\\
Hi  &  isothermal  &  $\cdots$                     & 0.01    & $\cdots$ \\
\enddata
\tablenotetext{a}{The halo gas fraction $f_{\rm hg} = M_{\rm hg}/M_{\rm h}$}
\tablenotetext{b}{The disk gas fraction $f_{\rm dg} = M_{\rm dg}/M_{\rm d}$}
\end{deluxetable*}

\begin{deluxetable*}{llcccc}
\tablecolumns{6}
\tablewidth{0pc}
\footnotesize
\tablecaption{Parameters of each component of the initial galaxy models \label{tab02}}
\tablehead{
\colhead{ } &
\colhead{ } &
\colhead{Model DHi\tablenotemark{a} /DHi-f5} &
\colhead{Model DHn/DHn-f5}&
\colhead{Model D} &
\colhead{Model Hi}
}
\startdata
\bf{Star disk:}\\
disk model & & exponential&  exponential & exponential & exponential \\

$a_{\rm ds}$ [kpc]\tablenotemark{b} & Length scale of star disk    & 3.5  & 3.5 & 3.5 & 3.5\\
$z_{\rm ds}$ [kpc]            & Vertical disk scale height   & 0.35 & 0.35 & 0.35 &  0.35\\
$b_{\rm ds}$ [kpc] & Outer disk cut-off radius & 42 &  42 & 42& 42\\
$M_{\rm ds}$ [$10^{10}$ M$_{\odot}$]  & Total mass of star disk & 4.4 & 4.4 & 4.4 &  5.0 \\
$N_{\rm ds}$       & Number of particles  & 16384 & 16384 & 16384&  16384\\
$m_{\rm ds}$ [$10^{10}$ M$_{\odot}$] & Mass of individual particles & $2.69 \times 10^{-4}$ &
                                              $2.69 \times 10^{-4}$ &
                                              $2.69 \times 10^{-4}$ &
                                              $3.05 \times 10^{-4}$ \\
$\epsilon_{\rm ds}$ [kpc] & Gravitational softening length  &
                             0.609/0.621 & 0.609/0.621  &  0.606 & 0.649 \\

\cline{1-6} \\
\bf{Gas disk:}\\
disk model & & exponential &  exponential & exponential & $\cdots$ \\

$a_{\rm dg}$ [kpc] & Length scale of gas disk     & 8.75  &   8.75 & 8.75 & $\cdots$ \\
$z_{\rm dg}$ [kpc] & Vertical disk scale height    & 0.35 &   0.35& 0.35 & $\cdots$\\
$b_{\rm dg}$ [kpc] & Outer disk cut-off radius & 105 &  105 & 105 & $\cdots$ \\
$M_{\rm dg}$ [$10^{10}$ M$_{\odot}$] & Total mass of gas disk   & 0.6 &  0.6 &  0.6 & $\cdots$ \\
$N_{\rm dg}$        & Number of particles  & 16384 &  16384 & 16384 &  $\cdots$\\
$m_{\rm dg}$ [$10^{10}$ M$_{\odot}$] & Mass of individual particles & $3.66 \times 10^{-5}$ &
                                              $3.66 \times 10^{-5}$ &
                                              $3.66 \times 10^{-5}$ &
                                              $\cdots$ \\
$\epsilon_{\rm dg}$ [kpc] & Gravitational softening length  &
                             0.225/0.229 & 0.225/0.229  &  0.224 & $\cdots$ \\
\cline{1-6} \\
\bf{DM halo:} \\
halo model & & NFW &  NFW  & NFW & NFW\\
$a_{\rm hd}$  [kpc]       & Radial scale of DM halo   &   21   & 21 & 21 & 21   \\
$b_{\rm hd}$ [kpc]        & Radius to begin tapering    & 84 & 84 & 84  & 84   \\
$M_{\rm hd}(a_{\rm hd})$ [$10^{10}$ M$_{\odot}$]  & Mass within radius $a_{\rm hd}$
         & 12.23/11.74  & 12.23/11.74 &  12.35 & 12.23 \\
$M_{\rm hd}(\infty) = M_{\rm hd}$  & Total mass of DM halo  &
        118.8/114  & 118.8/114  & 120 & 118.8 \\
$N_{\rm hd}$ & Number of particles
         & 163840 & 163840& 163840 & 163840\\
$m_{\rm hd}$ [$10^{10}$ M$_{\odot}$] & Mass of individual particles
                                            & $7.25 \times 10^{-4}$ / &
                                              $7.25 \times 10^{-4}$ / &
                                              $7.32 \times 10^{-4}$ &
                                              $7.25 \times 10^{-4}$ \\
 & &  $6.99 \times 10^{-4}$    &  $6.99 \times 10^{-4}$ \\
$\epsilon_{\rm hd}$ [kpc]  & Gravitational softening length  &
                             1 & 1 & 1 & 1  \\
\cline{1-6} \\
\bf{Gas halo:} \\
halo model & &  isothermal  & NFW &   $\cdots$ & isothermal \\
$a_{\rm hg}$ [kpc]   & Radial scale, or   &   $\cdots$ &    21  &  $\cdots$ & $\cdots$\\
              & $\,\,\,$ radius of core &  10.5   & $\cdots$ & $\cdots$ &  10.5 \\
$b_{\rm hg}$ [kpc]  & Radius to begin taper, or   & $\cdots$&  84  & $\cdots$ & $\cdots$\\
             & $\,\,\,$ radius of taper   & 420 & $\cdots$ &  $\cdots$ & 420 \\
$M_{\rm hg}(a_{\rm hg})$ [$10^{10}$ M$_{\odot}$] & Mass within radius $a_{\rm hg}$
                                & $\cdots$ & 0.12/0.62 &  $\cdots$ & $\cdots$\\
$M_{\rm hg}$ [$10^{10}$ M$_{\odot}$] & Total mass of gas halo
                & 1.2/6.0 &  1.2/6.0  & $\cdots$ & 1.2 \\
$N_{\rm hg}$ & Number of particles
         & 32768/163840  & 32768/163840 &  $\cdots$ & 32768\\
$m_{\rm hg}$  [$10^{10}$ M$_{\odot}$] & Mass of individual particles & $3.66 \times 10^{-5}$ &
                                              $3.66 \times 10^{-5}$ &
                                              $\cdots$ &
                                              $3.66 \times 10^{-5}$ \\
$\epsilon_{\rm hg}$ [kpc]  & Gravitational softening length  &
                             0.225/0.229 & 0.225/0.229  &  $\cdots$ & 0.225 \\
\cline{1-6} \\
\bf{Bulge:} \\
bulge model & &  Hernquist &    Hernquist &  Hernquist & Hernquist\\
$a_{\rm b}$ [kpc] & Length scale of bulge & 0.7 & 0.7 &  0.7 & 0.7  \\
$b_{\rm b}$ [kpc] & Radius at which truncation starts &  140 & 140& 140 & 140\\
$M_{\rm b}$ [$10^{10}$ M$_{\odot}$] & Total mass of bulge  & 1 & 1 &  1 & 1 \\
$N_{\rm b}$ & Number of particles
         & 8192 & 8192 & 8192 & 8192  \\
$m_{\rm b}$ [$10^{10}$ M$_{\odot}$] & Mass of individual particles & $1.22 \times 10^{-4}$ &
                                              $1.22 \times 10^{-4}$ &
                                              $1.22 \times 10^{-4}$ &
                                              $1.22 \times 10^{-4}$ \\
$\epsilon_{\rm b}$ [kpc] & Gravitational softening length  &
                             0.410/0.419 & 0.410/0.419  &  0.408 & 0.410 \\
\enddata
\tablenotetext{a}{Models~DHi and DHir adopt the same parameter values.}
\tablenotetext{b} {Physical units are used in this table,
$h^{-1}$~kpc for the length and $10^{10}$ M$_{\odot}$
for the mass parameter values.
Conversion from dimensionless code units is described in the text.}
\end{deluxetable*}

Our models are characterized by
the existence of a gas halo
and denoted by initials~DH, D, and H.
(D and H stand for disk and halo, respectively.):
The models of type~DH, such as the fiducial model~DHi,
consist of three collisionless components,
dark matter (DM) halo, stellar disk, and stellar bulge,
as well as two gaseous components, gas halo and gas disk.
The models of types~D and H also possess the three collisionless components
but have only one gaseous component at the initial time,
either gas disk in type~D or gas halo in type~H.

In all types of our models, the total mass
is set to
$M_{\rm tot}=126 \times 10^{10}\,\rm{M_{\odot}}$,
i.e., as similarly massive as the Milky Way Galaxy.
The total masses of the halo (DM + gas, if the model includes a gas halo),
the disk (stars + gas, if the model has a gas disk),
and the bulge (stars only) components are
$M_{\rm h}= M_{\rm hd} + M_{\rm hg} = 120 \times 10^{10}\,\rm{M_{\odot}}$,
$M_{\rm d}= M_{\rm ds} + M_{\rm dg} = 5 \times 10^{10}\,\rm{M_{\odot}}$,
and $M_{\rm b}=1 \times 10^{10}\,\rm{M_{\odot}}$, respectively in all models.

We test two different cumulative mass profiles for the gas halo,
the isothermal density profile (Eq.~3) and
the NFW profile (Eq.~2).
According to the adopted mass profile,
the models of type~DH are divided into two subtypes - DHi and DHn.
The gas halo of the models of subtype~DHi
(models~DHi, DHi-f5, and DHir) and model~Hi
follows the isothermal profile,
and that of subtype~DHn (models~DHn and DHn-f5) follows the NFW profile.

We also try two distinct values of halo gas fraction
$f_{\rm hg} = M_{\rm hg}/M_{\rm h}$ (Table~1),
in order to examine how galaxy models with different amounts of hot gas
evolve differently.
In one case,
the halo gas fraction in models~DHi, DHir, DHn, and Hi
(without `-f5' in the model names)
is chosen to be $f_{\rm hg} = 0.01$,
which corresponds to $M_{\rm hg} = 1.2 \times 10^{10}\,\rm{M_{\odot}}$.
This mass of a gaseous halo is motivated from Anderson \& Bregman (2010)
where the upper limit of the hot halo mass of the Milky Way Galaxy is estimated
to be $< 1.2-1.5 \times 10^{10}\,\rm{M_{\odot}}$
from the observed dispersion measure of pulsars in the Large Magellanic Cloud,
assuming an NFW profile for the hot gas.
In the other case, in models~DHi-f5 and DHn-f5,
a more massive gas halo with $f_{\rm hg} = 0.05$ is adopted,
which corresponds to $M_{\rm hg} = 6 \times 10^{10}\,\rm{M_{\odot}}$.
The more massive gas halo is motivated from \citealt{Moster2011}.
In their model with the maximum hot gas fraction,
the mass of the gas halo
within the virial radius is
$M_{\rm hg} = 1.2 \times 10^{11}\,\rm{M_{\odot}}$,
such that the baryonic fraction within the virial radius
has the universal value.

The gas fraction in the disk
is chosen to be a single value of $f_{\rm dg} = M_{\rm dg}/M_{\rm d}$ = 0.12
and thus $M_{\rm dg} = 0.6 \times 10^{10}\,\rm{M_{\odot}}$
in all models possessing a gas disk.
So the total mass of cold gas in a disk is either
half (in models DHi, DHir, and DHn) or a tenth (in models DHi-f5 and DHn-f5)
of the total mass of the hot halo gas.

The effects of the rotation of the hot gas halo is also examined
by comparing models DHi and DHir; the latter is made from the former
by adding an initial spin to the gas halo in such a way that
the gas halo has the rotation curve similar to that of the gas disk,
but with about half the amplitude (rotational velocity).
The hot gas particles in model~DHir thus have non-zero initial velocities in
$x$ and $y$ directions but zero in $z$ direction;
in all other models with a gas halo,
halo gas particles have zero initial velocities in all three directions.

The initial temperatures of the disk gas particles
in all models possessing a gas disk
are set to the single value of $10000~K$.
The temperatures
of the halo gas particles are determined
by the hydrostatic equilibrium in the ZENO code (Appendix~A).
The gas temperature in the central part of our models within $r = 10$~kpc
generally has a value between $10^6~K$ and $10^7~K$.

We summarize the key properties of each initial galaxy model below.


\subsubsection{Model DHi} 

Model~DHi is the fiducial model.
As noted earlier, it consists of
five components:
star disk, gas disk, DM halo, gas halo, and bulge.
The total mass of each of the components
in the unit of $10^{10}\,\rm{M_{\odot}}$ are
$M_{\rm ds} = 4.4$, $M_{\rm dg} = 0.6$,
$M_{\rm hd} = 118.8$, $M_{\rm hg} = 1.2$, and $M_{\rm b} = 1$
as summarized in Table~2.
Hence, the mass of the hot gas initialized in the halo is twice
that of the cold gas in the disk.
In terms of gas fractions in the disk and the halo,
$f_{\rm dg} = 0.12$ and $f_{\rm hg} = 0.01$ as indicated in Table~1.
The number of star or gas particles set in each component is
$N_{\rm ds} = 16384$, $N_{\rm dg} = 16384$,
$N_{\rm hd} = 163840$, $N_{\rm hg} = 32768$, and $N_{\rm b} = 8192$.
The number of gas particles in the halo is twice of that in the disk,
making the same mass per gas particle in the system.
The particle mass is different for different components.
To minimize the effects of the mass difference of simulation particles
we set $f_i \propto m_i/\epsilon_i^2$ to be equal
for particles in each component,
where $f_i$ is the maximum acceleration experienced by a particle.
We choose the gravitational softening length for the DM halo particles
to $\epsilon_{\rm hd}$ = 1~kpc (in all models; Table~2).
Then the smoothing lengths for the gas, disk star, and bulge particles
are determined to be 0.225, 0.609, and 0.410~kpc, respectively.

The initial distribution of the star and gas particles
and some basic profiles of the components
are shown in Figs~1 and 2, respectively.
As for both star and gas disks,
the particle distributions in the $x$-$y$ and $x$-$z$ projections
are presented in the first two columns of Fig.~1,
and their surface density profiles
in the top-left panel of Fig.~2.
Both disks have the exponential surface density profile (Eq.~1).
The gas disk is seen more extended than the stellar one,
with the larger radial scale length of
$a_{\rm dg} = 2.5 \times a_{\rm ds} = 8.75$~kpc (Table~2).
The particle distributions of the three spheroidal components
are shown in the third to fifth columns in Fig.~1,
and the spherically averaged accumulated density profiles
are plotted
in the top-right panel of Fig.~2.
The DM halo follows the NFW profile with the exponential taper (Eq.~2).
The radial scale is set to $a_{\rm hd} = 21$~kpc,
the same as in the galaxy model of \citet{McMillan2007},
and the radius to begin taper to $b_{\rm hd} = 84$~kpc.
The mass of the DM halo within $r = a_{\rm hd}$ is set to
$M_{\rm hd}(a_{\rm hd}) = 12.23 \times 10^{10}\,\rm{M_{\odot}}$
so that the total mass of the DM halo
becomes $M_{\rm hd} = 118.8 \times 10^{10}\,\rm{M_{\odot}}$.
The gas halo has the truncated isothermal profile (Eq.~3),
with the core radius $a_{\rm hg} = 10.5$~kpc
and the taper radius $b_{\rm hg} = 420$~kpc.
The choice of the core radius is motivated by the recent model of Moster et al. (2011).
Given the core radius,
the gas halo has nearly flat density within $a_{\rm hg}$,
as seen in the magenta line in the top-right panel of Fig.~2.
The bulge follows the Hernquist profile with a truncation (Eq.~4).
The length scale of the bulge is set to
$a_{\rm b} = 0.7$~kpc, adopted from McMillan and Dehnen (2007).

\begin{figure*}[!hbt]
\centering%
\includegraphics[width=14cm]{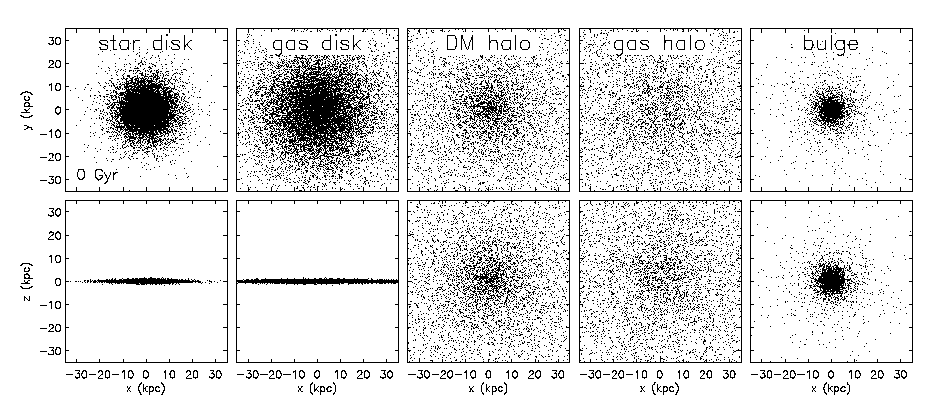}
\caption{Initial particle distribution of models~DHi and DHir.
(Both models have exactly the same spatial distribution of the particles,
but the latter has an additional spin of the gas halo.)
The star or gas particles of the five
components are displayed separately.
The top and bottom rows show the distribution projected
on to the $x$-$y$ and $x$-$z$ plane, respectively.
For the DM halo component, every tenth particle is plotted;
for the other components, every particle is plotted.
Other initial galaxy models have a similar distribution as model~DHi,
except the gas halo (see text for the details).}
\end{figure*}

\begin{figure*}[!hbt]
\centering%
\includegraphics[width=14cm]{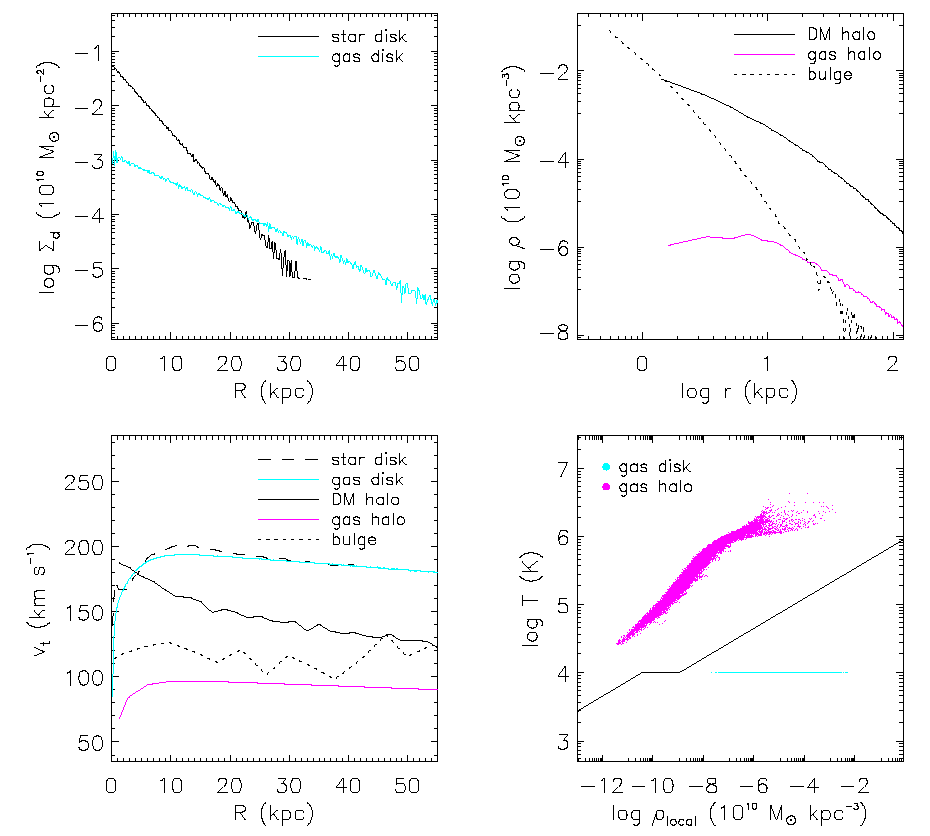}
\caption{Initial properties of models~DHi and DHir.
$Top~left$: Surface density profiles of the star and gas disk components.
(Those of the other initial galaxy models are shown in Fig.~4.)
$Top~right$: Spherically averaged accumulated density profiles
of the DM and gas halo and the bulge components.
(The gas halo profiles of other models are also shown in Fig.~4.)
$Bottom~left$: Cylindrically averaged tangential velocity profiles of
the star and gas disks, the DM and gas halos, and the bulge.
The gas halo of model~DHir only has the non-zero $v_{\rm t}$
(but zero in $v_z$)
as drawn with the magenta solid.
The gas halos of model~DHi and all the other models have
zero initial velocities in all three directions.
(The $v_{\rm t}$ profiles of the four components except gas halo
of the other models are similar as those shown in the panel.)
$Bottom~right$: Temperature against local density for the gas particles
in the disk and halo components.
The particles set as the gas disk and the gas halo are marked with
cyan and magenta dots, respectively.
The temperature of the disk gas particles is initialized
to a single value of $T = 10000$~K.
The temperature of the halo gas is determined
to achieve hydrodynamic equilibrium.
The disk and halo gas particles are initially well-separated
by the black line in the panel.
This line is chosen to use as a fixed criterion to divide the disk and halo gas
at later times.
}
\end{figure*}

The cylindrically averaged tangential velocities
$v_{\rm t} = (v_{x}^2 + v_{y}^2)^{1/2}$
of all components with respect to $R$ are shown
in the bottom-left panel of Fig.~2.
The gas disk particles are initially given
to have the local circular velocities (cyan solid line)
in clockwise direction
with zero vertical motion ($v_z = 0$),
the star disk particles to have additional dispersions (black dashed line).
Note that the gas halo particles of this model
are set with zero initial velocities in all three directions.

The temperature $T$
as a function of
the local gas density $\rho_{\rm local}$
in the disk or halo
is presented in the bottom-right panel of Fig.~2.
(The temperature is converted from the specific internal energy $u$ of the gas particles,
assuming that $T$ below 10000~K, a mean molecular weight
corresponding to neutral gas of primordial abundance,
otherwise full ionization.)
The temperature of the disk gas particles
are set uniformly to $T = 10000$~K (cyan dots).
The temperature of the halo gas
varies (magenta dots) to achieve hydrostatic equilibrium
of the hot halo gas.
Those halo gas particles
in the $T$-$\rho_{\rm local}$ diagram
appear well-separated
from the cold disk gas at the initial time.
As the system evolves (the model evolution is discussed in Sec.~3),
$T$ and $\rho_{\rm local}$ of the gas particles will be redistributed
and some of the halo and the disk particles will be mixed.
For example, some cold gas particles which are originally set as the disk gas
can be heated and have physical
characteristics of the halo gas, and vice versa.
Therefore, we establish a simple criterion,
shown by the black line in the $T$-$\rho_{\rm local}$ plane,
to determine if a gas particle at a certain time
should be considered as the disk or the halo particle.
The criterion is fixed in all our models at all times (c.f. Fig.~3).
The criterion is originally set in the $u$-$\rho_{\rm local}$ plane
and is converted on the assumption described above.
The line is chosen such that (1) the disk and the halo gas particles of
all initial galaxy models can be well-separated by the common line,
(2) all disk gas particles in the model without a gas halo
and without considering winds
can stay below the line at all times,
(3) and all halo gas particles in the model
without a gas disk and without winds can always stay above the line.
Any gas particles lie below the black line at a time
are considered as the disk gas
and those above the line as the halo gas particles,
with their origin identified by cyan or magenta
in the $T$-$\rho_{\rm local}$ plot.


\subsubsection{Model DHi-f5}

This model is intended to have a more massive gas halo
than that of the fiducial model~DHi,
with the other properties in common (Tables~1 and 2; to
keep the total mass of the DM + gas halo the same, the
DM halo of model~DHi-f5 is set to be lighter than that
of model~DHi by the mass difference in the gas halos).
Specifically, the gas halo of this model
follows the same truncated isothermal density profile of Eq.~3
as in the fiducial model
with the common values of $a_{\rm hg}$ and $b_{\rm hg}$,
but is five times more in mass $M_{\rm gh}$ as well as
in the number of particles $N_{\rm hg}$.
We keep the single (hot or cold) gas particle mass the same
at the initial time (in all our models).

As intended,
the components except the gas halo
in this model, and also in all other models of type~DH,
have (statistically) identical particle distributions at the initial time
to those of the fiducial model~DHi shown in Fig.~1.
In all models of type~DH,
the initial profiles of the four components,
such as the surface density profiles of both disks,
the spherically averaged density profiles of the DM halo and the bulge,
and the tangential velocity profiles of the four components,
are also the same (or nearly the same for the properties of the DM halo)
as those of the fiducial model presented in Fig.~2.
But the gas halo of this model has higher density at each radius
than that of the fiducial model~DHi
as shown in the second row-right column in Fig.~4
(see the black line for the initial density profile).
The initial velocities of the halo gas particles in this model
are set to zero as in the fiducial model.
The temperature $T$ with respect to the local gas density $\rho_{\rm local}$
for the gas particles in model~DHi-f5 is presented in Fig.~3.
As seen in the second row-leftmost column in the figure,
the disk gas particles are initialized to have a single value
of $T = 10000~K$ (cyan dots) as in the fiducial model
(and as also in all models with a gas disk),
but the temperature of the halo gas particles is assigned
in accordance with the hydrostatic equilibrium
of the halo gas (magenta dots).

\begin{figure*}[!hbt]
\centering%
\includegraphics[width=14cm]{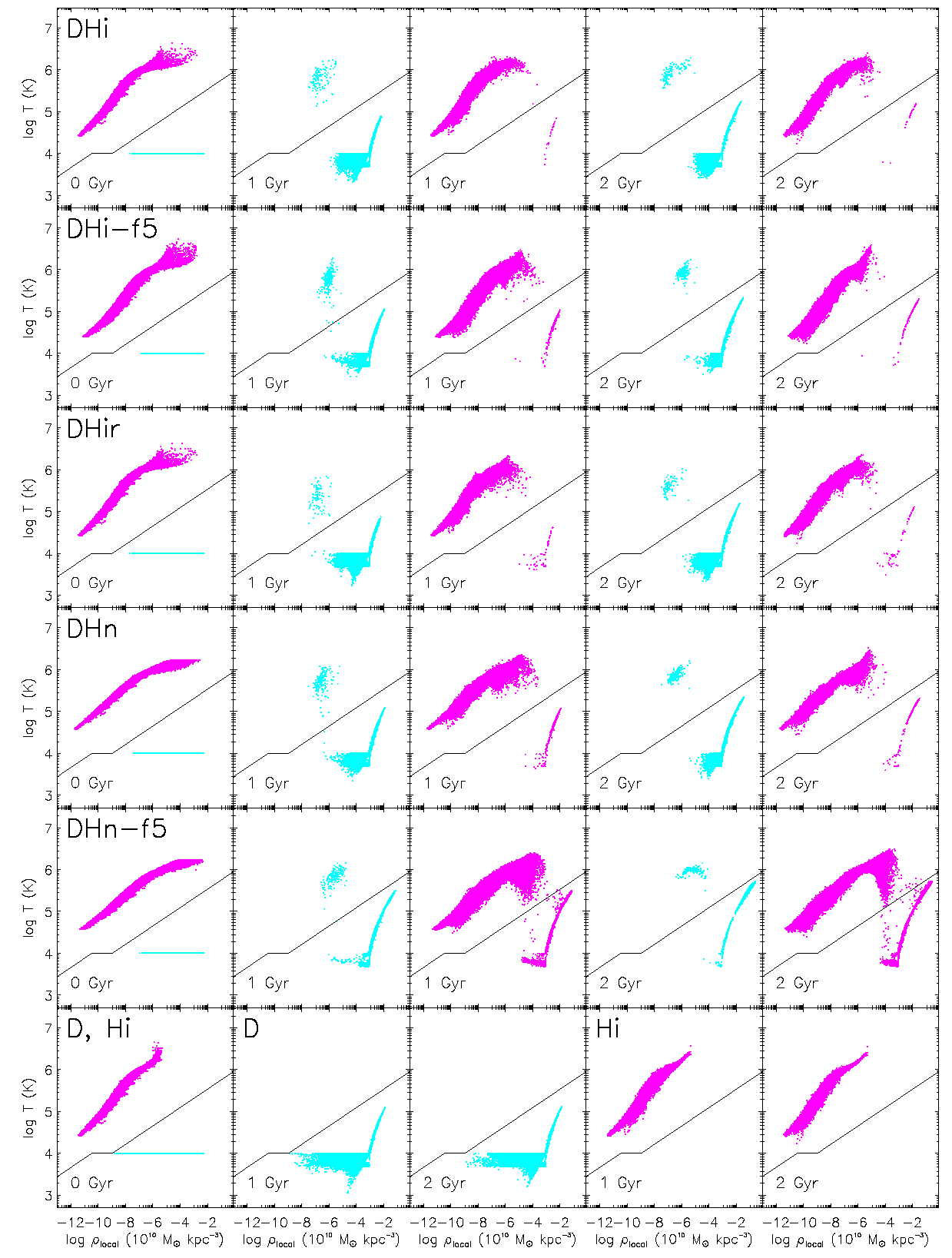}
\caption{
Temperature versus local density plots
for the disk and halo gas
of the galaxy models listed in Table~1.
The top through fifth rows show the plots from each model of type~DH.
The first column presents those at the initial time,
the second and third columns at $t$ = 1~Gyr,
and the fourth and fifth columns at $t$ = 2~Gyr.
The bottom row shows those from both models~D and Hi.
As same as the bottom-right panel of Fig.~2,
the gas particles originally set as the disk and the halo
are identified by cyan and magenta dots, respectively,
and the same black line is imposed in each panel
to divide the disk and halo gas at any time in all models.
Those cyan dots appear above the black line
are considered as halo gas at the time
together with the magenta dots remain above the line.
Similarly, those magenta dots lie below the line
are counted as disk gas together with the cyans remain below the line.
}
\end{figure*}

\begin{figure*}[!hbt]
\centering%
\includegraphics[width=14cm]{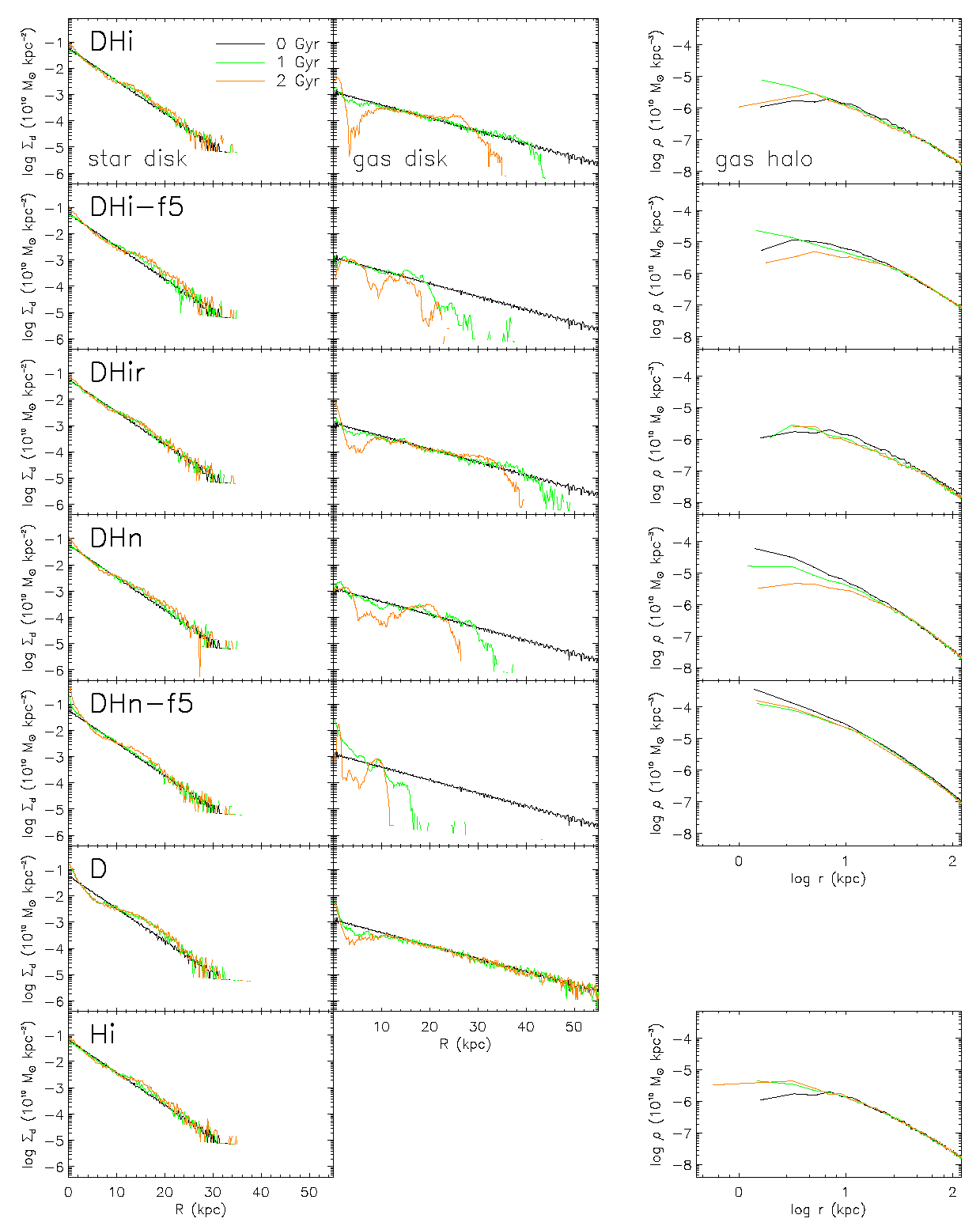}
\caption{The surface and spherically averaged density profiles
of the galaxy models at $t$ = 0, 1, and 2~Gyr.
The left and middle columns present the surface density profiles of the
star and gas disks, respectively,
and the right column the spherically averaged accumulated density profiles of
the gas halos.
Each row shows the profiles from each of the seven models.}
\end{figure*}


\subsubsection{Model DHir}

Model~DHir is designed to examine how the rotation of the gas halo
makes a difference in the evolution.
Thus the only difference between model~DHir and the fiducial model~DHi is
whether the gas halo is spinning with non-zero initial velocities (in model~DHir)
or stationary at the beginning.
Model~DHir shares all parameter values of all components given in Table~2
with the fiducial model.

The initial distribution of the particles and some basic profiles
are presented in Figs~1 and 2,
which are common to both models DHir and DHi
except the tangential velocity profile of the gas halo as designed
(bottom-left panel of Fig.~2).
The tangential velocities of the hot gas in model~DHir
are assigned in such a way that
the rotation curve of the gas halo
has the same shape but has half the amplitude of that of the gas disk.
The vertical velocity $v_z$ is set to zero in this model.
The choice of the gas halo rotation is not yet strongly restricted by current observations.
For example, \citet{Moster2011} chose to set the spin of a gas halo in such a way that
the specific angular momentum of the gas halo is four times that of the DM halo
(see Sec.~2 of their paper for details).
The test of different gas halo rotations would be the subject of future research.


\subsubsection{Model DHn}

As one of the DH type models,
model DHn consists of the five components of both star and gas disks,
both DM and gas halos, and a bulge.
Differently from the fiducial model~DHi,
the gas halo of this model follows the NFW profile
instead of the isothermal profile (Tables~1 and 2).
The initial total mass of the gas halo is
the same as that of the fiducial model.
This model is used to examine
how the different initial density profiles of the hot gas
with equal mass
make differences in the evolution.

The particle distributions in the four components except the gas halo
in model~DHn
are identical to those of the fiducial model in Fig.~1.
The gas halo follows the NFW profile like the DM halo
but has five times fewer particles than in the DM halo.
The spherically averaged density profile of the gas halo
is presented in Fig.~4 (fourth row-right column).
It is much higher
within a few tens of kpc
than that of the fiducial model
and monotonically increasing towards the center.
The initial tangential velocity of the gas halo is zero in this model.
The temperature $T$ of the gas particles
at the initial time
is presented in Fig.~3 with respect to the local density $\rho_{\rm local}$
(fourth row-leftmost panel).


\subsubsection{Model DHn-f5}

Model~DHn-f5, like the previous model~DHn,
possesses a gas halo following the NFW profile
but has five times more hot halo gas particles (Tables~1 and 2).

The spatial distribution of the hot gas
following the NFW profile has the same shape as
that of the DM halo shown in Fig.~1.
The number of the hot halo gas particles is
equal to that of the halo DM particles.
A halo gas particle has 1/19 times lower mass than
a DM particle.
The spherically averaged density profile of the gas halo
is specifically shown in Fig.~4 (fifth row-right column).
Model~DHn-f5 has the highest density at the inner part
among all our models.
The initial velocity of the gas halo
is set to zero.
The temperature of the gas as a function of
the local gas density is plotted
in Fig.~3 (fifth row-leftmost panel).


\subsubsection{Model D}

This model is composed of star and gas disks, a DM halo,
and a bulge, but has no gas halo.
The disk gas fraction and the total mass and the number of particles
of each component are summarized in Tables~1 and 2.

The initial particle distributions of the four components in this model
are equal to those of model~DHi (Fig.~1).
The initial $T$ versus $\rho_{\rm local}$ plot for the gas disk is
presented in Fig.~3 (bottom row-leftmost column; cyan dots only).


\subsubsection{Model Hi}

Model~Hi is similar to the fiducial model~DHi but has no gas disk.
Its gas halo follows the isothermal density profile (Tables~1 and 2).

The initial particle distributions of the four components
are equal to those of model~DHi (Fig.~1).
The spherically averaged initial density profile
of the hot halo
is shown in Fig.~4 (bottom row-right column; black line),
which is the same as that of model~DHi.
The initial velocity of the gas halo
is zero in this model.
The surface density of the star disk
(bottom row-left column in Fig.~4; black line) is highest among
those of all models,
as it has the more massive star disk without a gas disk.
The initial $T$ versus $\rho_{\rm local}$ diagram for the gas halo
is presented in Fig.~3 (bottom row-leftmost column; magenta dots only);
it is different from that of the fiducial model
due to the different galactic potentials of the models.


\section{SIMULATIONS}

We use an early version of GADGET-3 code (originally described in Springel 2005)
to evolve our galaxy models listed in Table~1
in isolation for 3~Gyr.
The simulation code
provides a model for galactic winds driven by star formation.
We first run the seven models without including galactic winds.
We then perform four more simulations using
two of our models - models~DHir and D -
including winds with different sets of wind parameter values
as summarized in Table~3.
We will refer to the four runs with winds
as `wind test runs' (Table~3) and name them
DHir-Wa, DHir-Wa-e1, DHir-Wi, and D-Wa.

In this section, we first describe the simulation code
to explain some key processes.
Then we present our simulation results.

\begin{table}
\begin{center}
\centering%
\caption{Wind test runs \label{tab03}}
\doublerulesep2.0pt
\renewcommand\arraystretch{1.5}
\begin{tabular}{lccc}
\hline \hline
Name     & Wind mode  & Wind efficiency ($\eta$)   \\
\hline
DHir-Wa    & axial      &  2      \\
DHir-Wa-e1 & axial      &  1      \\
DHir-Wi    & isotropic  &  2      \\
\hline
D-Wa       & axial      &  2      \\
\hline
\end{tabular}
\end{center}
\end{table}


\subsection{The Simulation Code}

The GADGET-3 code is a parallel TreeSPH code.
The gravitational force is computed
with a hierarchical multipole expansion
and the hydrodynamical force
with the smoothed particle hydrodynamics (SPH) technique
in the entropy conservative formulation \citep{Springel2002}.
The code adopts radiative cooling and heating by photoionization
\citep{Katz1996}.
It also includes star formation and supernova feedback
using the sub-resolution multiphase model of the interstellar medium (ISM)
developed by \citet{Springel2003}.
The ISM is pictured as a fluid comprised of condensed clouds
in pressure equilibrium with an ambient hot medium.
A thermal instability is assumed to be operating
in the two-phase region exceeding a threshold density $\rho_{\rm th}$.
Star formation occurs in the dense regions,
and converts cold clouds into stars
on a time-scale constrained by observations \citep{Kennicutt1998}.
The mass fraction of massive stars among the newly formed stars
is given by the parameter $\beta$.
The massive stars are subject to die
instantly as supernovae (with metal enrichment)
releasing energy to the ambient phase.
Some cold clouds are supposed to evaporate inside the supernova bubbles.
The evaporation of cold clouds inside of supernova bubbles
is handled by the simplified treatment
of \citeauthor{Springel2003}~(\citeyear{Springel2003}; see Sec.~3 of
their paper for details).
Only the density dependency on the supernova evaporation is taken into account
for simplicity \citep{Mckee1977}.
The radiative cooling of the hot gas causes
a corresponding increase of the cold gas.
These processes - star formation, cloud evaporation and cloud growth -
lead to a self-regulated cycle for star formation.
The multiphase model is numerically implemented
with the simplified treatment described in Sec.~5.2 in \citet{Springel2003}
based on the fact that the conditions of the self-regulation are achieved quickly.
In the treatment, the mass of each new star particle is set identically
to $m_{\star} = m_{\rm 0}/N_{\rm g}$,
where $m_{\rm 0}$ is the initial gas mass before the star formation
and $N_{\rm g}$ is number of star particles that may be generated
in a gas particle at each star formation event.

The code also includes a model for
galactic winds driven by star formation.
The rate of the disk mass loss through a wind
is assumed to be proportional to the star formation rate
$\dot{M}_{\rm w} = \eta \dot{M}_{\star}$,
where $\eta$ is a coefficient measuring the wind efficiency
and $\dot{M}_{\star}$ is the formation rate of long-lived stars.
It is also assumed that
the rate of energy loss through the wind is
$\dot{M}_{\rm w} v_{\rm w}^{2} /2 = \chi \epsilon_{\rm SN} \dot{M}_{\star}$,
which is a fraction $\chi$ of the rate of energy generation by supernovae.
Here $v_{\rm w}$ is the velocity of the wind,
and $\epsilon_{\rm SN} = u_{\rm SN} \beta / (1-\beta)$
represents the average return of the supernova energy
from the stars formed following the \citet{Salpeter1955} initial mass function.
The velocity of a particle
is modified to
$\vec{v^{\prime}} = \vec{v} + v_{\rm w} \hat{n}$
by the wind.
The wind can be axial or isotropic \citep{Springel2003};
the direction of the unit vector $\hat{n}$ is chosen to be either
random on the unit sphere in isotropic winds
or as a random orientation along the direction of
$\vec{v} \times \nabla \phi$ ($\phi$ is the gravitational potential)
in axial winds.

In all of our simulations,
the standard parameters for the multiphase model are used.
They are mostly the same as those in \citet{Springel2003},
otherwise adjusted based on later observations or findings.
The star formation time-scale
$t_{0}^{\star}$ is set to 1.47~Gyr, and
the mass fraction of massive star to $\beta$ = 0.1.
The `supernova temperature' $T_{\rm SN} = 2 \mu u_{\rm SN}/(3k)$
is chosen to be $10^{8}$~K which is equivalent to
$\epsilon_{\rm SN} = 4 \times 10^{48}$~erg/$\rm{M_{\odot}}$,
the temperature of cold clouds $T_{\rm c}$ to 1000~K,
and the parameter for supernova evaporation $A_{0}$ to 1000.
We adopt $N_{\rm g}$ = 2,
so a new star particle
has a half the gas particle mass before the star formation event.
Correspondingly, the total number of particles increases
due to star formation
even though the total mass of a system is kept constant.

In the wind test runs (Table~3),
the effects of galactic winds driven by supernovae are included.
Axial winds are tested in runs DHir-Wa, DHir-Wa-e1, and D-Wa,
and isotropic winds in run DHir-Wi.
The parameter of the wind efficiency is set to the standard value $\eta$ = 2
in runs DHir-Wa, DHir-Wi, and D-Wa,
while the half the value in DHir-Wa-e1 for comparison.
The wind energy fraction parameter is set to
the default value $\chi$ = 1 in the runs with winds.


\subsection{The Evolution of Galaxy Models Without Winds}

Here we describe the evolution of our galaxy models without winds
(listed in Table~1), focusing on the effects of the hot gas halo.

The distributions of particles in each simulation at four epochs
are presented in Figs~5 through 11.
We present the snapshots of the runs
together with the evolution of the $T$-$\rho_{\rm local}$ plots (Fig.~3),
the density profiles of the star disk, the gas disk, and the gas halo (Fig.~4),
and the specific mass and the particle numbers of the components (Table~4).
We begin with the runs of type~DHi models.

\begin{figure*}[!hbt]
\centering%
\includegraphics[width=14cm]{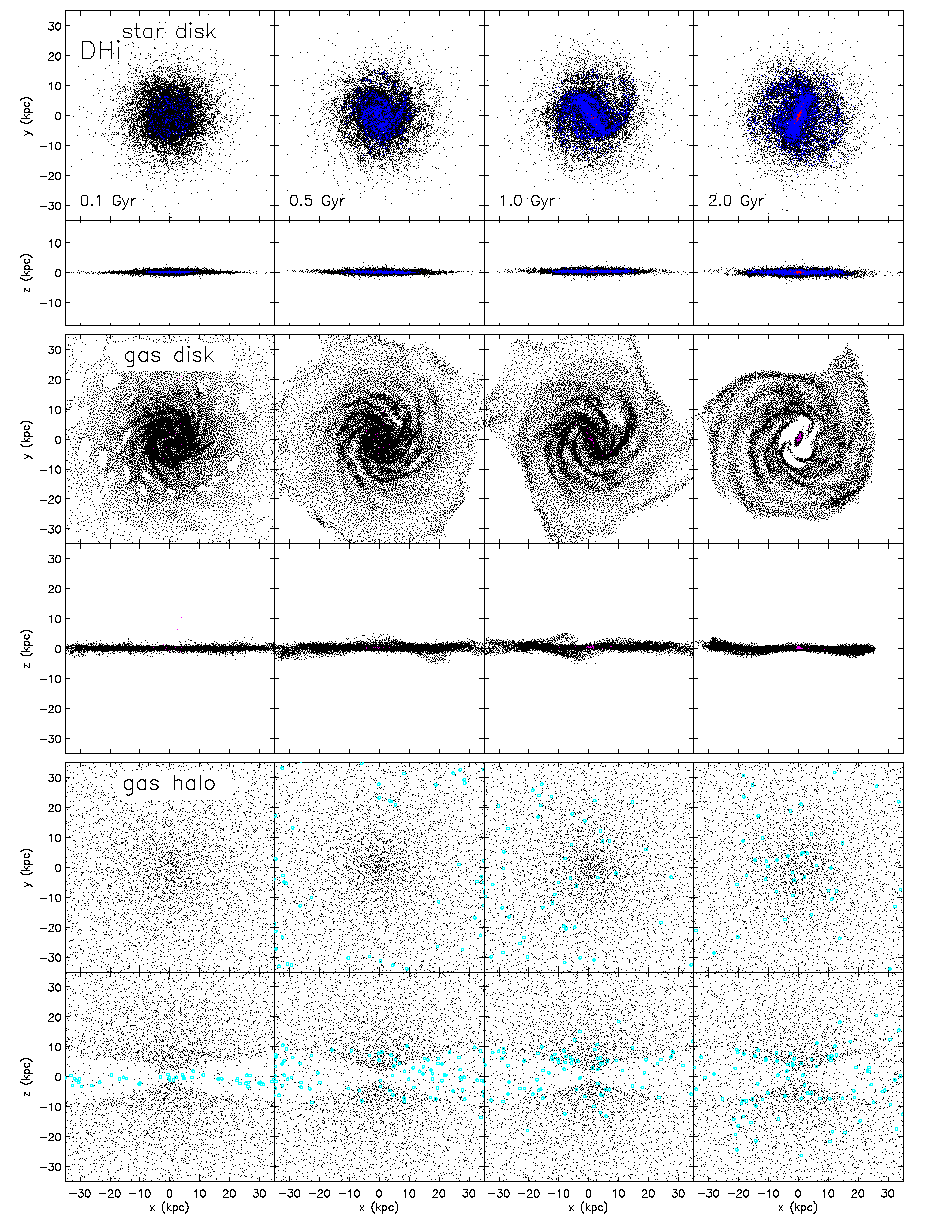}
\caption{Four snapshots of the distribution of particles in run~DHi.
The first to fourth columns represent the system
at $t$ = 0.1, 0.5, 1, and 2~Gyr, respectively.
The first two rows show the distribution of the star particles in the disk
seen in the $x$-$y$ plane (first row) and the $x$-$z$ plane (second row).
The particles originally set as the star disk are marked with black dots and
the particles turned into stars originating from the gas disk and the gas halo
are indicated with blue and red dots, respectively.
The third and fourth rows show the gas disk in the two projections.
The particles originally set as the gas disk are marked in black dots and
those accreted from the gas halo are in magenta dots.
The bottom two rows show the particles in the gas halo.
The particles originally set as the gas halo are marked in black dots and
those included to the halo from originally the gas disk are in cyan circles.
The cyan markers are bigger than the others to better distinguish
the particles.
}
\end{figure*}

\begin{figure*}[!hbt]
\centering%
\includegraphics[width=14cm]{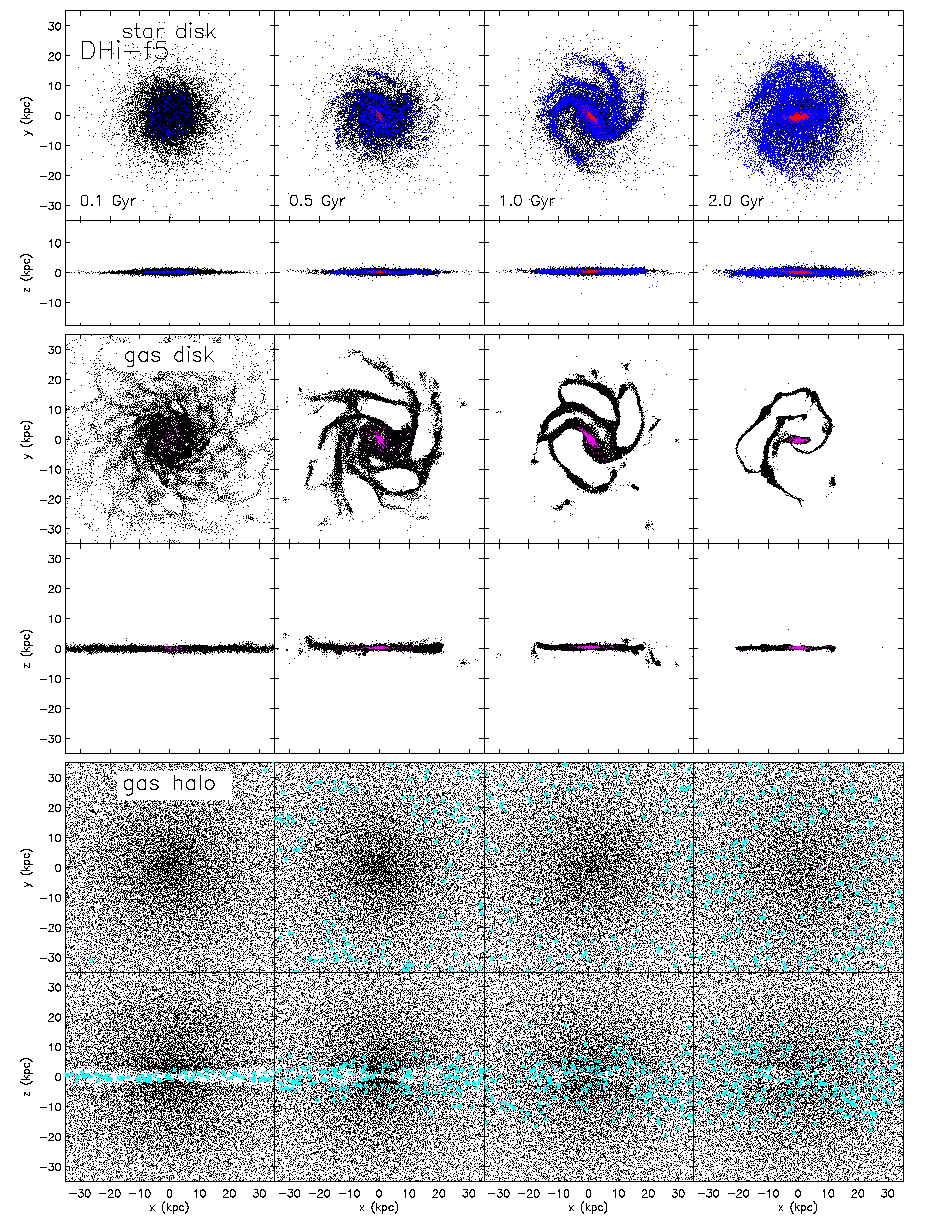}
\caption{The same snapshots as Fig.~5, but from run~DHi-f5.
}
\end{figure*}

\begin{figure*}[!hbt]
\centering%
\includegraphics[width=14cm]{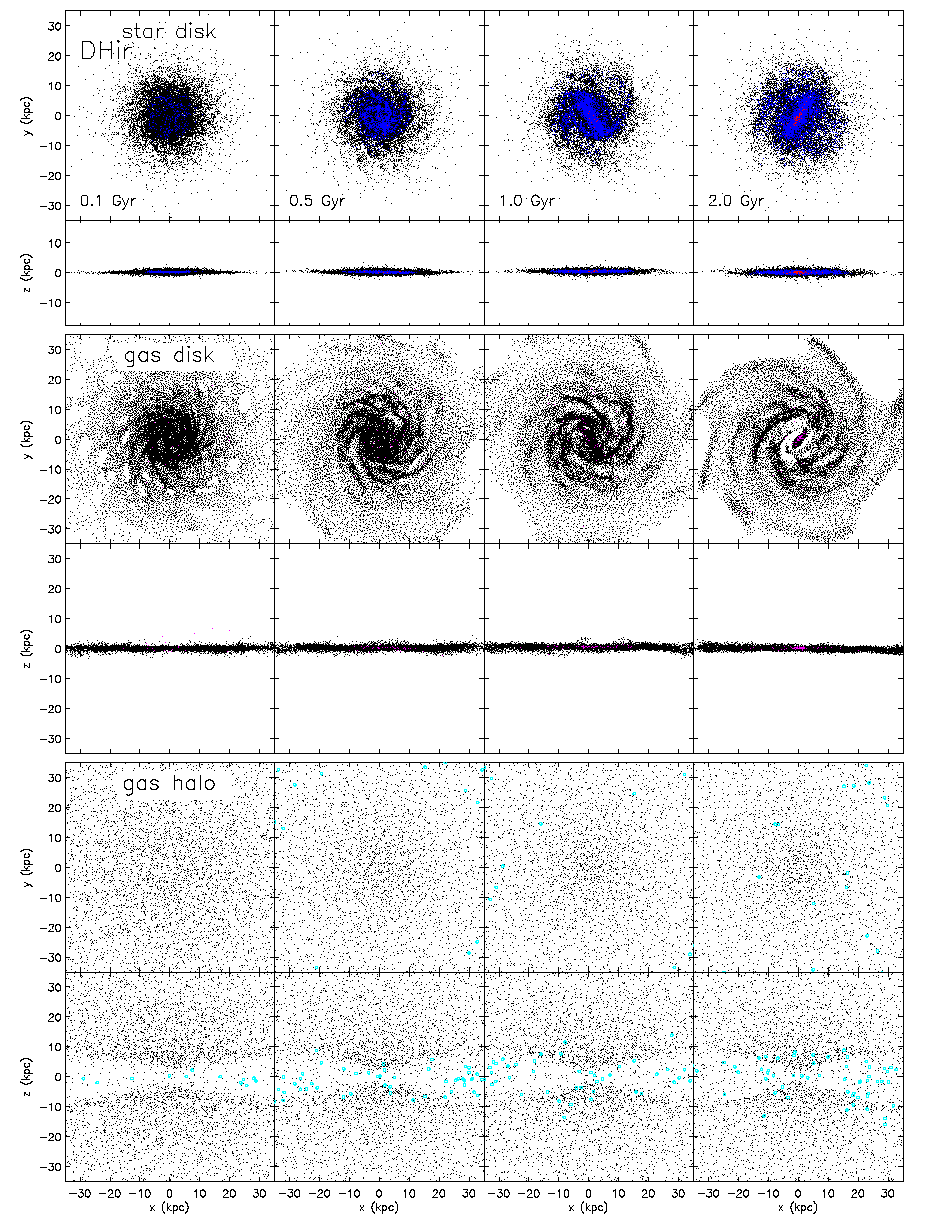}
\caption{The same snapshots as Fig.~5, but from run~DHir.
}
\end{figure*}

\begin{figure*}[!hbt]
\centering%
\includegraphics[width=14cm]{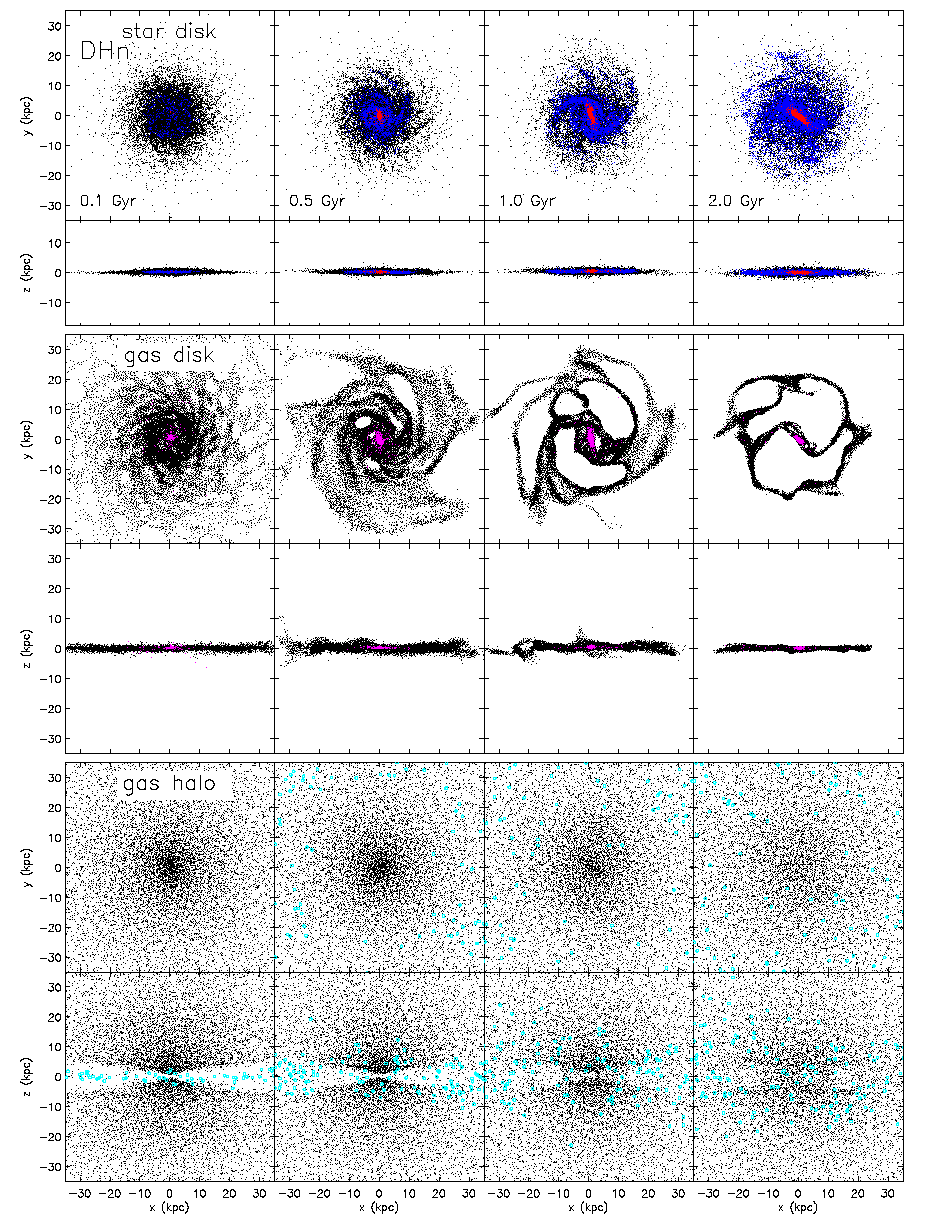}
\caption{The same snapshots as Fig.~5, but from run~DHn.
}
\end{figure*}

\begin{figure*}[!hbt]
\centering%
\includegraphics[width=14cm]{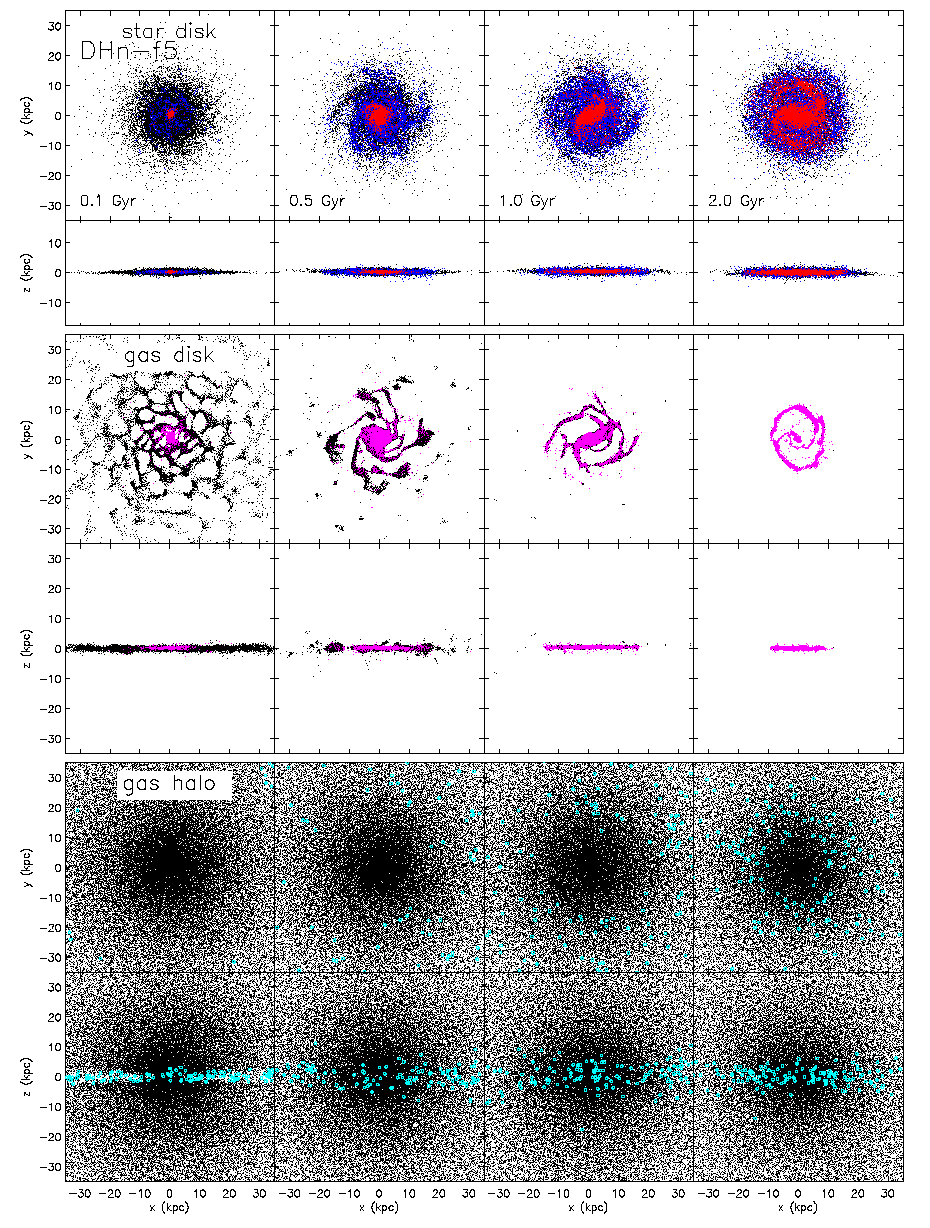}
\caption{The same snapshots as Fig.~5, but from run~DHn-f5.
}
\end{figure*}

\begin{figure*}[!hbt]
\centering%
\includegraphics[width=14cm]{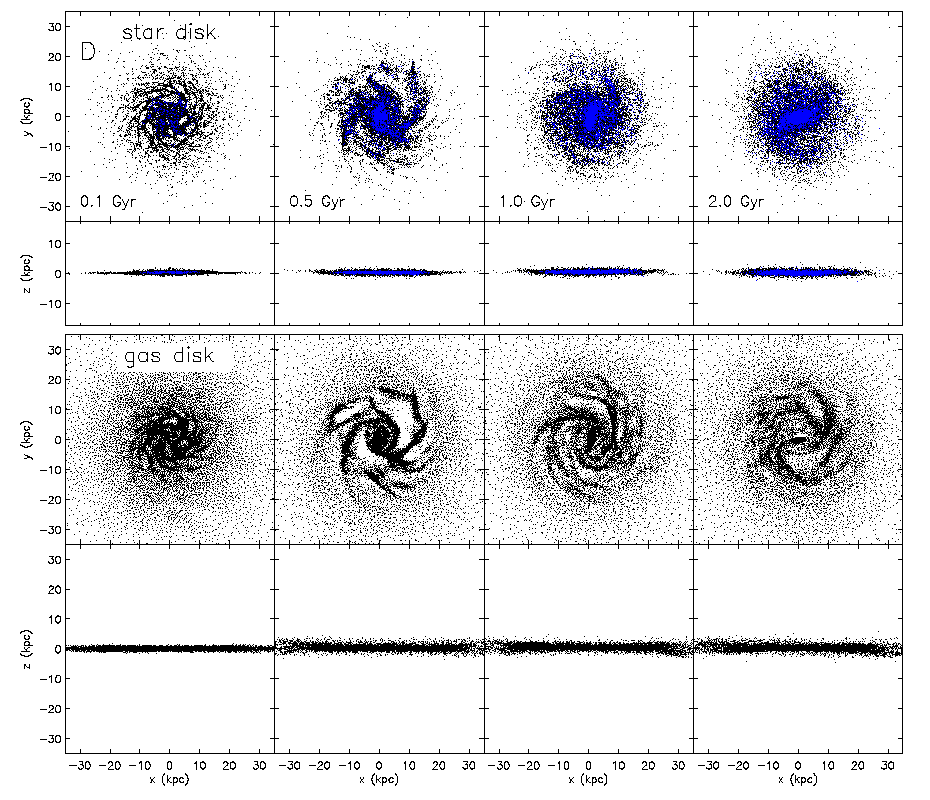}
\caption{The same snapshots as Fig.~5, but from run~D.
The initial galaxy model~D does not have a gas halo.}
\end{figure*}

\begin{figure*}[!hbt]
\centering%
\includegraphics[width=14cm]{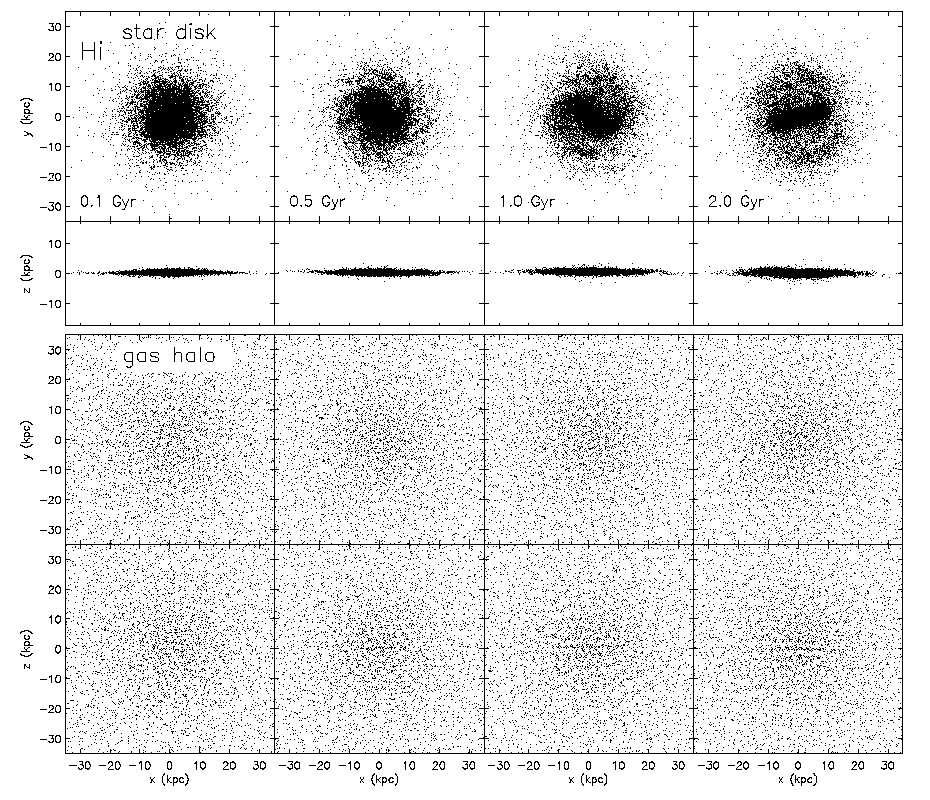}
\caption{The same snapshots as Fig.~5, but from run~Hi.
The initial galaxy model~Hi does not have a gas disk.}
\end{figure*}

\begin{deluxetable*}{ l |rrrr|rrr|rrr  }
\tablecolumns{11}
\tablewidth{0pc}
\tiny
\tablecaption{Mass and number of particles of the star and gas disks, and the gas halo
in all of our simulations. \label{tab04}}

\startdata
\cline{1-11} \\[0.1mm]
\cline{1-11} \\[0.5mm]
  &
$N_{\rm ds}$ ($M_{\rm ds}$)\tablenotemark{a} &
$N_{\rm ds,org}$ &
$N_{\rm ds,d2s}$ &
$N_{\rm ds,h2s}$ &
$N_{\rm dg}$ ($M_{\rm dg}$) &
$N_{\rm dg,org}$ &
$N_{\rm dg,h2d}$ &
$N_{\rm hg}$ ($M_{\rm hg}$) &
$N_{\rm hg,org}$ &
$N_{\rm hg,d2h}$\\

\cline{1-11} \\[0.5mm]

\bf{DHi} &&& &&& &&&\\
$t$ = 0~Gyr &
16384 (4.399)  &   16384 &  0 &  0 &
16384 (0.6000) &   16384 &  0 &
32768 (1.200)   &  32768 &  0 \\
$t$ = 0.1~Gyr &
16780 (4.407)  &   16384 &    396 &  0 &
16270 (0.5888)   &  16260 &    10 &
32877 (1.204)    &  32758 &   119\\
$t$ = 0.5~Gyr &
18165 (4.432)    &   16384  &   1780 &   1 &
16127 (0.5629)   &   16115  &   12  &
32890 (1.204)    &   32756  &   134 \\
$t$ = 1~Gyr &
19678 (4.460)   &   16384  &  3286 &  8 &
15794 (0.5359)  &   15771  &  23   &
32871 (1.203)   &   32745  &  126 \\
$t$ = 2~Gyr &
23182 (4.524)   &   16384 &   6756 &  42 &
14089 (0.4726)  &   14057 &   32 &
32849 (1.203)   &   32724 &   125 \\
&&& &&& &&&\\
\bf{DHi-f5} &&& &&& &&&\\
$t$ = 0~Gyr &
16384 (4.399)   &   16384  &  0 &  0 &
16384 (0.6000)   &  16384  &  0 &
163840 (5.998)  &   163840 &  0 \\
$t$ = 0.1~Gyr &
16851 (4.408)   &  16384 &  465 &   2 &
16190 (0.5848)   &  16149 &    41 &
164022 (6.004)   &  163799 &   223 \\
$t$ = 0.5~Gyr &
19244 (4.452)   &   16384 &  2813 &   47 &
16074 (0.5450)  &  15904 &  170 &
163912 (6.000)   &  163668 &  244 \\
$t$ = 1~Gyr &
24232 (4.543)   &   16384  &  7624 &   224 &
15200 (0.4598)   &  14911  &  289 &
163745 (5.994)   &   163505 &  240 \\
$t$ = 2~Gyr &
36350 (4.765)   &   16384  &  19290 &   676 &
9795 (0.2411)    &   9648  &  147 &
163658 (5.991)   &   163416 & 242\\
&&& &&& &&&\\
\bf{DHir} &&& &&& &&&\\
$t$ = 0~Gyr &
16384 (4.399)   &   16384 &  0 &  0 &
16384 (0.6000)  &  16384 &  0 &
32768 (1.200)   &   32768 &  0 \\
$t$ = 0.1~Gyr &
16724 (4.405)   &   16384 &   340 &  0 &
16363 (0.5932)  &  16344 &   19 &
32784 (1.200)   &   32749 &    35 \\
$t$ = 0.5~Gyr &
18079 (4.430)   &   16384 &   1689 &   6 &
16212 (0.5670)  &  16172 &  40 &
32823 (1.202)   &   32728 &  95 \\
$t$ = 1~Gyr &
19532 (4.457)   &   16384 &   3134 &  14 &
15903 (0.5410)  &  15850 &  53 &
32805 (1.201)   &   32713 &  92 \\
$t$ = 2~Gyr &
22621 (4.513)   &   16384 &   6173 &   64 &
14452 (0.4856)  &  14389 &   63 &
32775 (1.200)   &  32687 &   88 \\
&&& &&& &&&\\
\bf{DHn} &&& &&& &&&\\
$t$ = 0~Gyr  &
16384 (4.399)   &   16384 &  0 &  0 &
16384 (0.6000)  &  16384 &  0 &
32768 (1.200)   &  32768 &  0 \\
$t$ = 0.1~Gyr &
16860 (4.408)   &   16384 &  473 &   3 &
16378 (0.5915)  &  16218 &  160 &
32763 (1.199)   &  32608 &  155 \\
$t$ = 0.5~Gyr &
18752 (4.443)   &  16384 &   2173 &   195 &
16438 (0.5665)  &  16015 &   423 &
32499 (1.190)   &   32318 &  181 \\
$t$ = 1~Gyr &
21935 (4.501)   &   16384 &  4977 &  574 &
15862 (0.5164)  &  15318 &   544 &
32276 (1.182)   &   32087 &  189 \\
$t$ = 2~Gyr &
30479 (4.657)   &   16384 &   12743 &   1352 &
12060 (0.3646)  &  11832 &   228 &
32148 (1.177)   &   31955 &   193 \\
&&& &&& &&&\\
\bf{DHn-f5} &&& &&& &&&\\
$t$ = 0~Gyr &
16384 (4.399)    &   16384  &  0 &  0 &
16384 (0.6000)   &  16384  &  0 &
163840 (5.998)   &   163840 &  0 \\
$t$ = 0.1~Gyr &
17419 (4.418)   &   16384 &   849 &   186 &
17443 (0.6214)  &  16077 &   1366 &
162739 (5.957)   &  162462 &  277 \\
$t$ = 0.5~Gyr &
27734 (4.607)   &   16384 &  7821 &   3529 &
19764 (0.5906)  &  14926 &  4838 &
158422 (5.799)   &  158242 &  180 \\
$t$ = 1~Gyr &
43271 (4.892)    &   16384  &    16762 &  10125 &
17101 (0.4384)   &  11364  &    5737 &
154810 (5.667)   &   154646 &  164 \\
$t$ = 2~Gyr &
75100 (5.474)    &    16384  &    31139 &  27577 &
8256  (0.2252)    &  1213   &    7043 &
144716 (5.298)   &    144570 &  146 \\
&&& &&& &&&\\
\bf{D} &&& &&& &&&\\
$t$ = 0~Gyr &
16384 (4.399)   &   16384 & 0 & $\cdots$ &
16384 (0.6000)  &  16384 & $\cdots$ &
0 (0.0)         & $\cdots$        &  0 \\
$t$ = 0.1~Gyr &
16761 (4.406)   &   16384 &   377 & $\cdots$  &
16376 (0.5931)  &  16376 & $\cdots$  &
0  (0.0)     & $\cdots$      & 0 \\
$t$ = 0.5~Gyr &
18839 (4.444)   &   16384 &   2455 & $\cdots$  &
16150 (0.5551)  &  16150 & $\cdots$  &
0 (0.0)      & $\cdots$      &  0 \\
$t$ = 1~Gyr &
20767 (4.480)   &   16384 &  4383 & $\cdots$ &
15507 (0.5198)  &  15507 & $\cdots$ &
0 (0.0)         & $\cdots$          &  0 \\
$t$ = 2~Gyr &
22860 (4.518)   &   16384 &  6476 & $\cdots$ &
14342 (0.4815)  &  14342 & $\cdots$ &
0 (0.0)         & $\cdots$          &  0 \\
&&& &&& &&&\\
\bf{Hi} &&& &&& &&&\\
$t$ = 0~Gyr &
16384 (5.000)  &  16384 & $\cdots$  &  0 &
0 (0.0)        & $\cdots$        &  0 &
32768 (1.200)  &  32768 & $\cdots$  \\
$t$ = 0.1~Gyr &
16384 (5.000)   &  16384 & $\cdots$  &  0 &
0 (0.0)       & $\cdots$      &  0 &
32768 (1.200)  &   32768 & $\cdots$  \\
$t$ = 0.5~Gyr &
16384 (5.000)   &  16384 & $\cdots$  &  0 &
0 (0.0)       & $\cdots$      &  0 &
32768 (1.200)  &  32768 & $\cdots$  \\
$t$ = 1~Gyr &
16384 (5.000)  &  16384 & $\cdots$  &  0 &
0 (0.0)       & $\cdots$      &  0 &
32768 (1.200)  &  32768 & $\cdots$  \\
$t$ = 2~Gyr &
16384 (5.000)  &  16384 & $\cdots$  &  0 &
0 (0.0)       & $\cdots$      &  0 &
32768 (1.200)  &  32768 & $\cdots$  \\
&&& &&& &&&\\
\cline{1-11} &&& &&& &&&\\[0.5mm]

\bf{DHir-Wa} &&& &&& &&&\\
$t$ = 0~Gyr &
16384 (4.399)   &  16384 &  0 &  0 &
16384 (0.6000)  &  16384 &  0 &
32768 (1.200)   &   32768 &  0 \\
$t$ = 0.1~Gyr &
16720 (4.405)   &   16384 &  336 &  0 &
16054 (0.5823)  &  16040 &  14 &
33089 (1.211)   &   32754 &   335 \\
$t$ = 0.5~Gyr &
17521 (4.420)   &   16384 &   1137 &  0 &
15022 (0.5338)  &  15004 &  18 &
34071 (1.245)   &   32750 &   1321 \\
$t$ = 1~Gyr &
18137 (4.431)   &   16384 &  1753 &  0 &
14184 (0.4985)  &  14162 &  22 &
34817 (1.269)   &   32746 &  2071 \\
$t$ = 2~Gyr &
19124 (4.449)   &   16384 &    2737 &  3 &
12594 (0.4441)  &  12575 &  19 &
36102 (1.305)   &   32749 &     3353 \\
&&& &&& &&&\\
\bf{DHir-Wa-e1} &&& &&& &&&\\
$t$ = 0~Gyr &
16384 (4.399)   &   16384 &  0 &  0 &
16384 (0.6000)  &  16384 &  0 &
32768 (1.200)   &   32768 &  0 \\
$t$ = 0.1~Gyr &
16699 (4.405)   &   16384 &  315 &  0 &
16195 (0.5877)  &  16179 &  16 &
32951 (1.206)   &   32752 &   199 \\
$t$ = 0.5~Gyr &
17783 (4.425)   &   16384 &   1398 &  1 &
15505 (0.5469)  &  15484 &  21 &
33564 (1.227)   &   32747 &   817 \\
$t$ = 1~Gyr &
18642 (4.441)   &   16384 &   2256 &  2 &
14830 (0.5150)  &  14806 &   24 &
34080 (1.243)   &   32744 &   1336 \\
$t$ = 2~Gyr &
20218 (4.469)   &   16384 &   3828 &  6 &
13168 (0.4580)  &  13145 &  23 &
35134 (1.271)   &   32743 &   2391 \\
&&& &&& &&&\\
\bf{DHir-Wi} &&& &&& &&&\\
$t$ = 0~Gyr &
16384 (4.399)   &   16384 &  0 &  0 &
16384 (0.6000)  &  16384 &  0 &
32768 (1.200)   &   32768 &  0 \\
$t$ = 0.1~Gyr &
16752 (4.406)   &   16384 &  368 &  0 &
16117 (0.5840)  &  16110 &  7 &
33026 (1.209)   &   32761 &   265 \\
$t$ = 0.5~Gyr &
17609 (4.422)   &   16384 &   1224 &   1 &
14993 (0.5314)  &  14985 &  8 &
34095 (1.246)   &   32760 &   1335 \\
$t$ = 1~Gyr &
18268 (4.434)   &   16384 &   1883 &  1 &
14164 (0.4958)  &  14156 &  8 &
34822 (1.269)   &  32760 &   2062 \\
$t$ = 2~Gyr &
19434 (4.455)   &   16384 &  3047 &   3 &
12495 (0.4369)  &  12489 &   6 &
36151  (1.307)  &   32761 &  3390 \\
&&& &&& &&&\\
\bf{D-Wa} &&& &&& &&&\\
$t$ = 0~Gyr &
16384 (4.399)   &   16384 &  0 & $\cdots$ &
16384 (0.6000)  &  16384 & $\cdots$ &
0 (0.0)       & $\cdots$        &  0 \\
$t$ = 0.1~Gyr &
16716 (4.405)   &   16384 &   332 & $\cdots$ &
16355 (0.5932)  &  16355 & $\cdots$ &
21 (0.0007507)      & $\cdots$     &   21 \\
$t$ = 0.5~Gyr &
18193 (4.432)   &   16384 &   1809 & $\cdots$ &
15395 (0.5387)  &  15395 & $\cdots$ &
849 (0.02823)   & $\cdots$     &   849 \\
$t$ = 1~Gyr &
18947 (4.446)   &   16384 &  2563 & $\cdots$ &
14435 (0.5019)   &  14435 & $\cdots$ &
1629 (0.05125)   & $\cdots$        &  1629 \\
$t$ = 2~Gyr &
19417 (4.455)   &   16384 &  3033 & $\cdots$ &
13808 (0.4824)  &  13808 & $\cdots$ &
2071 (0.06211)  & $\cdots$        &  2071
\vspace{2mm}
\enddata
\tablenotetext{a}{All masses in this table are in units of $10^{10}$ M$_{\odot}$.}

\end{deluxetable*}


\subsubsection{Models having an isothermal gas halo}

Here we describe the results
for the galaxy models~DHi (the fiducial model), DHi-f5, and DHir.
These models contain a gas halo following the isothermal density profile (Eq.~3).

The snapshots of run~DHi are presented in Fig.~5.
The first two rows show the particle distribution of the star disk
projected in the $x$-$y$ and $x$-$z$ planes,
the third and fourth rows the gas disk,
and the bottom two rows the gas halo,
at $t$ = 0.1, 0.5, 1, and 2~Gyr
from the first to the fourth column sequentially.

In the first two rows in Fig.~5,
the black dots represent the stars originally set as the star disk,
and the blue and red dots indicate the stars added into the star disk
from the gas originating from (i.e. initially set as)
the gas disk and gas halo, respectively.
In Table~4 we give the total mass and
number of star disk particles ($M_{\rm ds}$ and $N_{\rm ds}$, respectively)
at each time;
we also specify the number of star disk particles that
are originally star disk particles ($N_{\rm ds,org}$),
produced from the disk gas ($N_{\rm ds,d2s}$),
and transformed from the halo gas ($N_{\rm ds,h2s}$) separately.
The total mass of the star disk as a whole is
$M_{\rm ds}$ $= M_{\rm ds,org} + M_{\rm ds,d2s} + M_{\rm ds,h2s}$.
Similarly, the total number of the star disk particles is
$N_{\rm ds} = N_{\rm ds,org} + N_{\rm ds,d2s} + N_{\rm ds,h2s}$.
As time passes, spiral patterns and a central bar develop in the disk.
Some deviation from the initial exponential surface density profiles
are seen in Fig.~4 (first row-left column) due to those structures generated.
The star disk grows in mass continuously (Table~4),
as some gas
turns into disk stars.
The mass increases from the initial value
$M_{\rm ds}$ = 4.399 $\times$ $10^{10}\,\rm{M_{\odot}}$
to 4.524 $\times$ $10^{10}\,\rm{M_{\odot}}$ at $t$ = 2~Gyr.
The total number of star disk particles also increases
from $N_{\rm ds}$ = 16384 at the beginning to 23182 at $t$ = 2~Gyr.

The disk gas particles shown in the middle two rows in Fig.~5
are those determined as the disk gas by the criterion
in the $T$-$\rho_{\rm local}$ plane (first row in Fig.~3).
The black dots in the snapshots are the original disk gas particles
still remaining in the disk
(i.e. in the $T$-$\rho_{\rm local}$ plot, the gas particles among cyan dots
lie below the criterion line).
The magenta dots superimposed on the black dots
are the originally hot halo particles that become
the disk gas according to the
$T$-$\rho_{\rm local}$ criterion
(i.e. the gas particles among magenta lie below the criterion line).
In the $T$-$\rho_{\rm local}$ diagrams at later times
(second to fifth panels of the first row in Fig.~3),
the disk gas particles, which are located below the black line,
follow the finger-like pattern in the higher density region
and show the sharp cut-off at $T \simeq 10000$~K in the lower density region.
The negative correlation between $T$ and $\rho_{\rm local}$
as shown by the finger-like feature results from
the effective equation of state for
star-forming gas in the multiphase model \citep{Springel2003}.
The sharp cut-off appears because the minimum temperature
a gas particle can reach through atomic radiative cooling processes
is about 10000~K.
As shown in the snapshots,
some spiral patterns develop in the gas disk as in the star disk;
warps are also generated as seen in the side view.
The gas disk dissipates as time passes.
Although some halo gas particles
cool down to become the gas disk particles
(mostly to the central part of the gas disk, as indicated with magenta dots),
more gas particles turn into the disk stars.
Some heated disk gas join the halo.
The number of particles that are initially disk gas particles ($N_{\rm dg,org}$)
and that converted from the halo gas ($N_{\rm dg,h2d}$)
at each epoch are given in Table~4.
As seen in the surface density profile of the gas disk
(first row-middle column in Fig.~4),
the size of the disk decreases by dissipation
and the profile also evolves.
The density at the innermost part of the disk increases at later times.
The total mass of the gas disk
$M_{\rm dg} = M_{\rm dg,org} + M_{\rm dg,h2d}$
decreases from the initial value
$M_{\rm dg}$ = 0.6 $\times$ $10^{10}\,\rm{M_{\odot}}$
to 0.4726 $\times$ $10^{10}\,\rm{M_{\odot}}$ after 2~Gyr (Table~4).

The black dots
in the bottom two rows in Fig.~5 are
the gas particles that remain as the halo gas from the beginning
(i.e. in the $T$-$\rho_{\rm local}$ diagram,
the gas particles among magenta dots lie above the criterion line).
The number of those particles is denoted by $N_{\rm hg,org}$ in Table~4.
The cyan circles superimposed on the black dots are those particles
among originally set as the cold disk gas particles but
having the properties of the gas halo
according to the $T$-$\rho_{\rm local}$ criterion
(i.e. the gas particle among cyan lie above the criterion line).
The number of these transformed halo gas particles is
$N_{\rm hg,d2h}$ in Table~4.
The cyan circles at early times
are located mostly at
the outer part of the gas disk.
At later times, they appear more dispersed in both $x$-$y$ and $x$-$z$ views.
The spherically averaged density profile of the gas halo is
shown in Fig.~4 (first row-right column).
At the initial time, the gas halo follows the isothermal density profile
with the halo gas fraction $f_{\rm hg}$ = 0.01 (black line).
It changes a bit (see olive and orange lines) from the initial profile
mainly in the central part,
as some halo gas are removed down to the disk
(which subsequently turn into disk stars) and the gas particles are redistributed.
The total mass of the gas halo $M_{\rm hg} = M_{\rm hg,org} + M_{\rm hg,d2h}$
changes only a little from
the initial value 1.2 $\times$ $10^{10}\,\rm{M_{\odot}}$
to 1.203 $\times$ $10^{10}\,\rm{M_{\odot}}$ at $t$ = 2~Gyr.

Fig.~6 shows the evolution of model~DHi-f5.
The model starts with
five times more hot gas ($f_{\rm hg} = 0.05$) than the fiducial model~DHi.
More blue and red dots are seen in the first two rows,
compared to the previous run.
Due to the massive gas halo,
the gas disk particles (originating from the gas disk or the gas halo)
are compressed, cooled, and turn into stars more quickly.
Consequently, the surface density profile of the star disk
(second row-left column in Fig.~4)
deviates from the initial exponential profile
more than that seen in run~DHi.
The star disk grows in mass from the initial value
$M_{\rm ds}$ = 4.399 $\times$ $10^{10}\,\rm{M_{\odot}}$ to
4.765 $\times$ $10^{10}\,\rm{M_{\odot}}$ at $t$ = 2~Gyr (Table~4).

The gas disk and the gas halo particles of run~DHi-f5
plotted in Fig.~6
are those determined in accordance with
the $T$-$\rho_{\rm local}$ criterion.
The gas disk dissipates very quickly as clearly seen in
the third and fourth rows in Fig.~6
and also in the surface density profile (second row-middle column in Fig.~4).
The gas disk diminishes in mass by star formation
from the initial value
$M_{\rm ds}$ = 0.6 $\times$ $10^{10}\,\rm{M_{\odot}}$ to
0.2411 $\times$ $10^{10}\,\rm{M_{\odot}}$ at $t$ = 2~Gyr (Table~4).
The gas halo changes mostly within a few tens of kpc as seen in
both the snapshots (Fig.~6) and the density profile (Fig.~4).
The change is contributed by the halo gas particles in the central part
which cool down to the disk (magenta dots in the snapshots)
as well as the gas particles originating from the gas disk
which are heated by the hot gas and moved to the more diffuse gas halo
(cyan circles in the snapshots).
Overall, the gas halo changes in mass slightly from the initial value of
$M_{\rm ds}$ = 5.998 $\times$ $10^{10}\,\rm{M_{\odot}}$ to
5.991 $\times$ $10^{10}\,\rm{M_{\odot}}$ at $t$ = 2~Gyr (Table~4).

Four snapshots of run~DHir are shown in Fig.~7.
The model has a rotating gas halo
following the isothermal density profile with $f_{\rm hg} = 0.01$.
Compared to the first run~DHi,
fewer number of gas particles originating from the gas disk
turn into stars (blue dots in the snapshots;
also refer to $N_{\rm ds,d2s}$ in Table~4).
On the other hand, more gas particles originating from the hot halo
cool down to the gas disk (magenta dots; $N_{\rm dg,h2d}$).
The accreted gas particles
mostly appear in the inner part of the disk but rarely in the outer part.
The rotation of the hot halo contributes the gas halo particles
to settle down to the disk a bit more efficiently,
but it also make the cooling time of the compressed gas particles in the disk longer.
This results in less massive star disk ($M_{\rm ds}$)
and more massive gas disk ($M_{\rm dg}$)
than those of run~DHi
(and also than all the runs of types DH and D without winds)
as shown in Table~4.
The gas disk in this run thus dissipates more slowly
than the two previous runs
as seen in both the snapshots and its surface density profile (Fig.~4).
Fewer number of gas particles originating from the disk
are heated and included into the halo (cyan circles; $N_{\rm hg,d2h}$).
The gas halo keeps its initial mass ($M_{\rm hg}$)
for at least a couple of Gyr (Table~4).


\subsubsection{Models having an NFW-profile gas halo}

Models~DHn and DHn-f5
have a gas halo following the NFW density profile (Eq. 2) with
$f_{\rm hg} = 0.01$ and 0.05, respectively.

Fig.~8 shows the four snapshots
from a run of DHn.
Compared to the run with the fiducial model~DHi,
more stars are added into the disk
from the disk and the halo gas
(blue and red dots; see also Table~4).
This is led by the higher density of the inner part of the gas halo
in model~DHn,
as seen in the spherically averaged density profiles (Fig.~4),
although the initial halo gas fraction
and thus the initial total mass of the hot gas
are identical in both models.
Compared to the run of DHi-f5 having a heavier gas halo,
fewer disk gas particles
turn into the stars
(blue dots; except at $t$ = 0.1~Gyr as shown in Table~4).
This is in part because the less massive gas halo in run~DHn
exerts overall weaker compression to the gas disk.
Note that, however, more halo gas particles
become the disk gas (magenta dots) in run~DHn compared to run~DHi-f5
and subsequently turns into stars (red dots),
due to the higher central density of the gas halo (Fig.~4),
even though the initial total mass of the gas halo
is five times less than that of model~DHi-f5.
In this run~DHn, the total mass of the star disk increases
from $M_{\rm ds}$ = 4.399 $\times$ $10^{10}\,\rm{M_{\odot}}$ at the initial time
to 4.657 $\times$ $10^{10}\,\rm{M_{\odot}}$ at $t$ = 2~Gyr (Table~4).
The gas disk and the gas halo
decrease in mass from
$M_{\rm dg}$ = 0.6 $\times$ $10^{10}\,\rm{M_{\odot}}$ at $t$ = 0
to 0.3646 $\times$ $10^{10}\,\rm{M_{\odot}}$ after 2~Gyr,
and from $M_{\rm hg}$ = 1.2 $\times$ $10^{10}\,\rm{M_{\odot}}$ at $t$ = 0
to 1.177 $\times$ $10^{10}\,\rm{M_{\odot}}$ after 2~Gyr, respectively.

Fig.~9 shows the the four snapshots from
the simulation of model~DHn-f5.
Its gas halo has the highest central density
among all our models (Fig.~4).
Star formation in the disk (blue and red dots in Fig.~9)
occurs most actively in this model run,
ending up with the most massive star disk of
$M_{\rm ds}$ = 5.474 $\times$ $10^{10}\,\rm{M_{\odot}}$ at $t$ = 2~Gyr
which is 124\% of its initial mass (Table~4).
The surface density of the star disk in the central part becomes
higher than all the other runs (Fig.~4).
The gas disk also dissipates very quickly in this run.
Most of the original disk gas
has been consumed by $t$ = 2~Gyr
(black dots in the middle two rows in Fig.~9; $N_{\rm dg,org}$ in Table~4);
the gas particles seen in the disk at later times are mostly those accreted
from the hot halo (magenta dots; $N_{\rm dg,h2d}$).
The total mass of the gas disk decreases to
$M_{\rm dg}$ = 0.2252 $\times$ $10^{10}\,\rm{M_{\odot}}$ at $t$ = 2~Gyr,
which is 38\% of its initial mass (Table~4).
The total mass of the gas halo also decreases
to 5.298 $\times$ $10^{10}\,\rm{M_{\odot}}$ at $t$ = 2~Gyr,
88\% of the initial mass.


\subsubsection{Models~D and Hi}

The four snapshots from run~D is presented in Fig.~10.
The galaxy model~D does not have a gas halo at the set-up.
New stars are added into the disk
as some gas in the disk turn into stars (blue dots).
None of the gas particles initially set as the disk gas
become the halo gas according to
the $T$-$\rho_{\rm local}$ criterion throughout the simulation
(left three panels of the bottom row in Fig.~3).
The star disk grows in mass from the initial value of
$M_{\rm ds}$ = 4.399 $\times$ $10^{10}\,\rm{M_{\odot}}$
to 4.518 $\times$ $10^{10}\,\rm{M_{\odot}}$ at $t$ = 2~Gyr (Table~4).
The gas disk mass $M_{\rm dg}$ decreases by the same amount
through star formation.
The radial extent of the gas disk
is maintained well compared to
all runs of type~DH models which possess a gas halo,
as seen in the snapshots (Fig.~10) and in the surface density profile (Fig.~4).

Fig.~11 shows the snapshots from run~Hi.
This model is initialized with the gas halo component
following the isothermal density profile
with the halo gas fraction $f_{\rm hg} = 0.01$
but does not have a gas disk.
Over the course of the simulation,
no gas particles initially set as the hot halo gas
cool down to the disk according to the $T$-$\rho_{\rm local}$ criterion
(leftmost panel and two rightmost panels in the bottom row in Fig.~3).
Consequently, no new stars are formed in the star disk;
the masses of the star disk and the gas halo remain
unchanged from the initial values (Table~4).
The surface density profile of the star disk deviates from
the initial exponential profile
due to the central bar and the spiral patterns developed in the star disk,
as seen in Fig.~4.
It is also seen that the spherically averaged density profile
of the gas halo increases
in the central part.


\subsection{The Evolution of Galaxy Models With Winds}

The evolution of the four wind test runs listed in Table~3
are discussed here.
As noted earlier, galactic winds driven by supernova are included
in these runs.

The temperature $T$ of the gas components
in the models with respect to the local density $\rho_{\rm local}$
are shown in Fig.~12,
and the evolution of the density profiles
of the star and gas disks and the gas halo in Fig.~13.
The particle distributions of the three components
in each run at four epochs are presented in Figs~14$-$17.
The total mass and number of particles of the components are given
in Table~4.

\begin{figure*}[!hbt]
\centering%
\includegraphics[width=14cm]{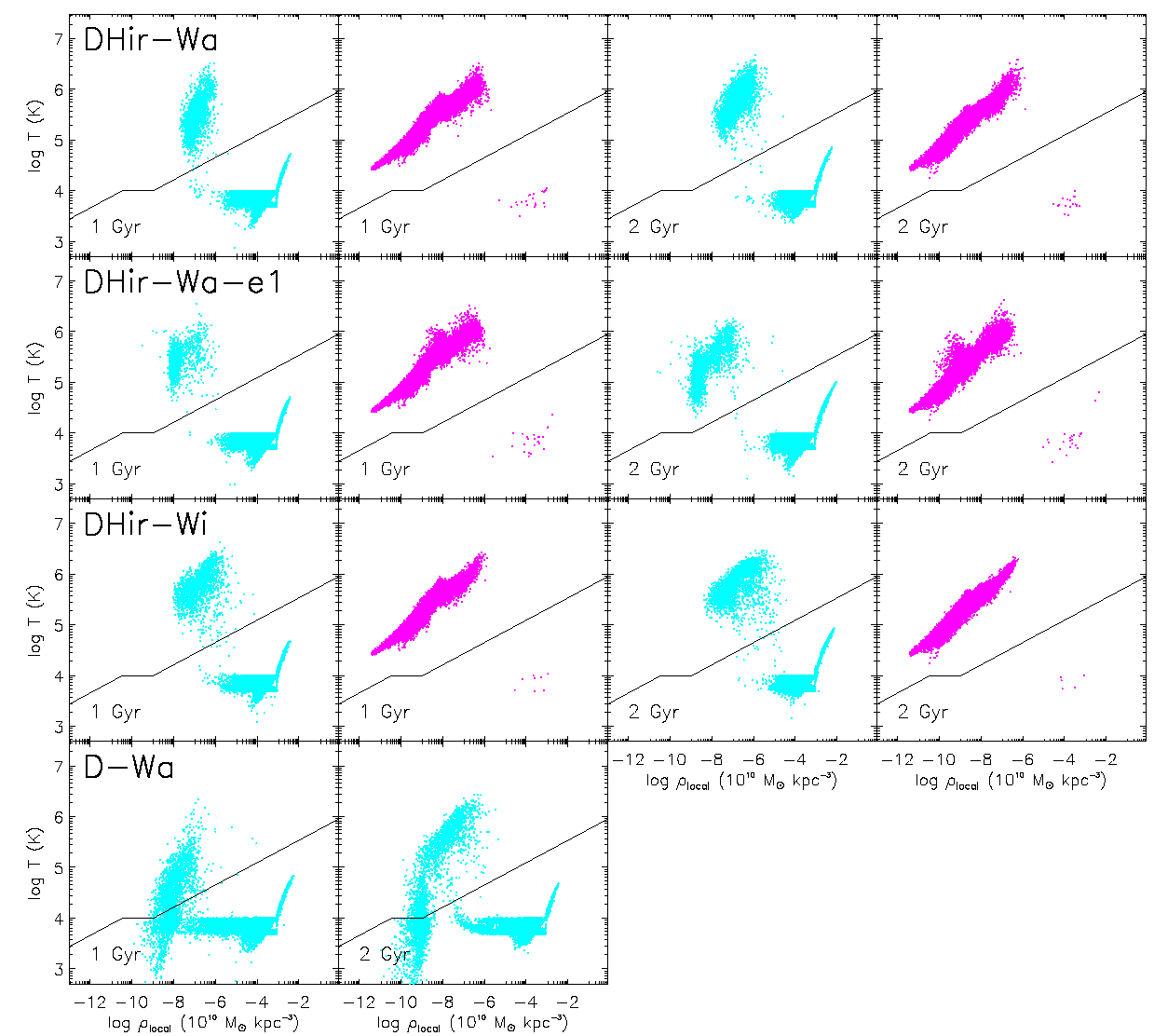}
\caption{
Temperature versus local density plots
for the disk and halo gas
of the wind test runs listed in Table~3.
As same as Fig.~3, the cyan and magenta dots indicate the gas particles
originally from the disk and halo, respectively, and the
identical black line is used to distinguish the disk and halo gas at later times.
Each row presents the plots from each run at $t$ = 1 and 2~Gyr.
The plots at the initial time are not shown here since
they are the same as those of either model~DHir or model~D in Fig.~3.
More gas particles crossing the black line upwardly are seen due to the wind effects
compared to the runs without winds.}
\end{figure*}

\begin{figure*}[!hbt]
\centering%
\includegraphics[width=14cm]{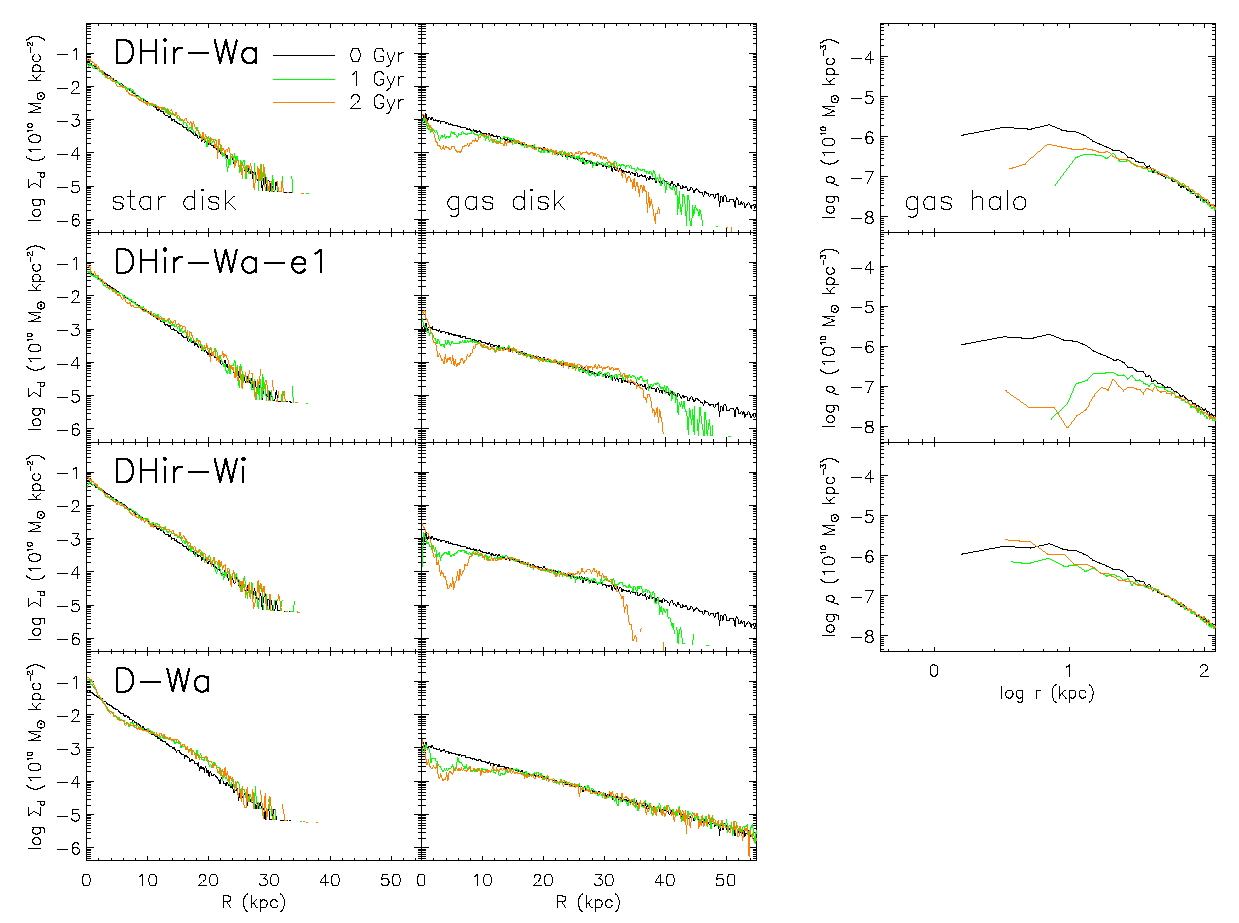}
\caption{
The surface and spherically averaged density profiles
of the wind test runs at $t$ = 0, 1, and 2~Gyr.
As same as Fig.~4, the left to right columns present
the surface density profiles of the star and gas disks,
and the spherically averaged accumulated density profiles of
the gas halos.
Each row shows the profiles from each run.
}
\end{figure*}

\begin{figure*}[!hbt]
\centering%
\includegraphics[width=14cm]{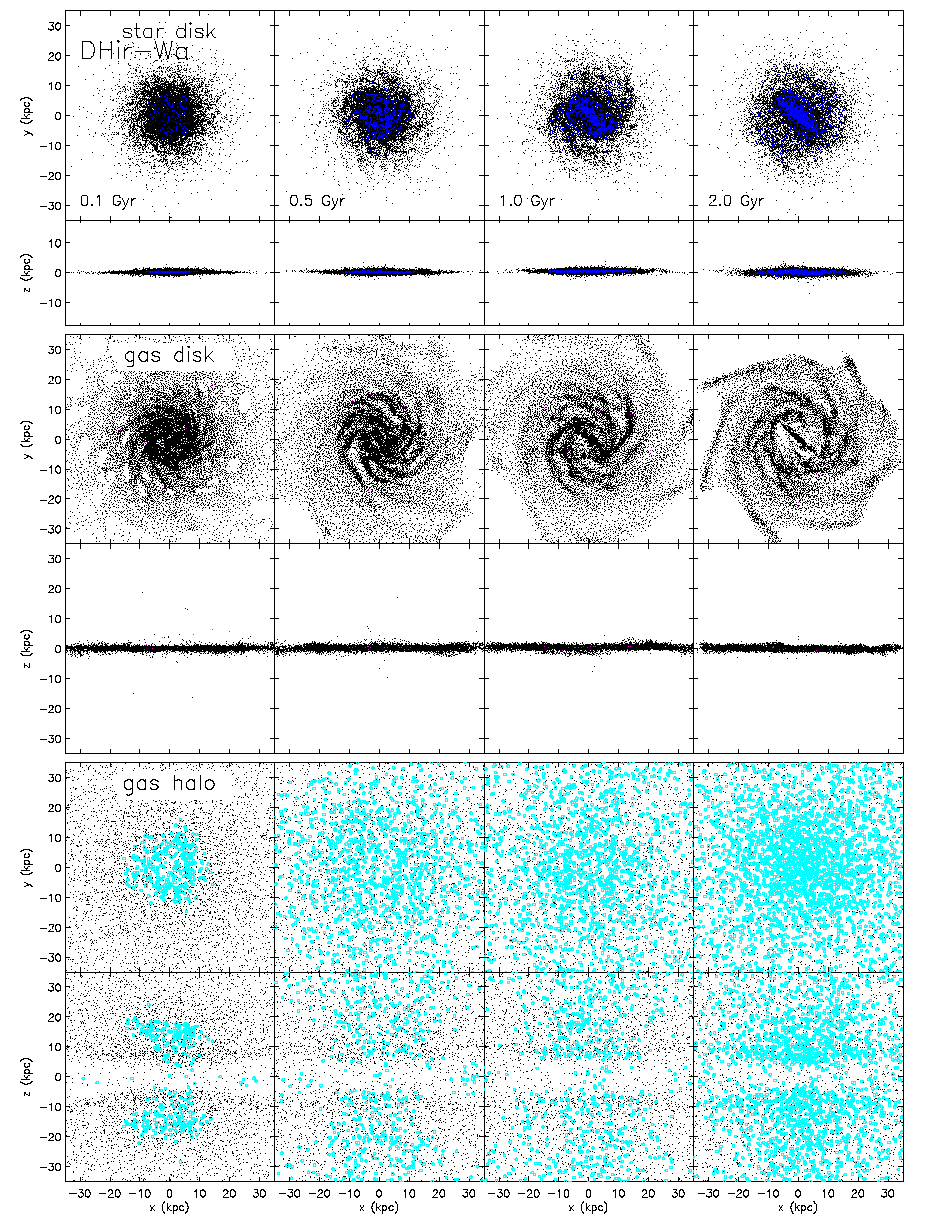}
\caption{
Four snapshots from run DHir-Wa which includes
axial galactic winds associated with star formation during the simulation.
These snapshots are taken at the same times as those in Figs~5 through~11
and the colors are also used in the same way.
Compared to the runs without winds,
the star disk tends to have less number of newly formed stars (blue and red dots).
Some gas particles are ejected from the plane of the disk by the stellar winds
out to several tens of kpc;
those ejected and heated gas are seen in the bottom two rows (cyan circles).
}
\end{figure*}

\begin{figure*}[!hbt]
\centering%
\includegraphics[width=14cm]{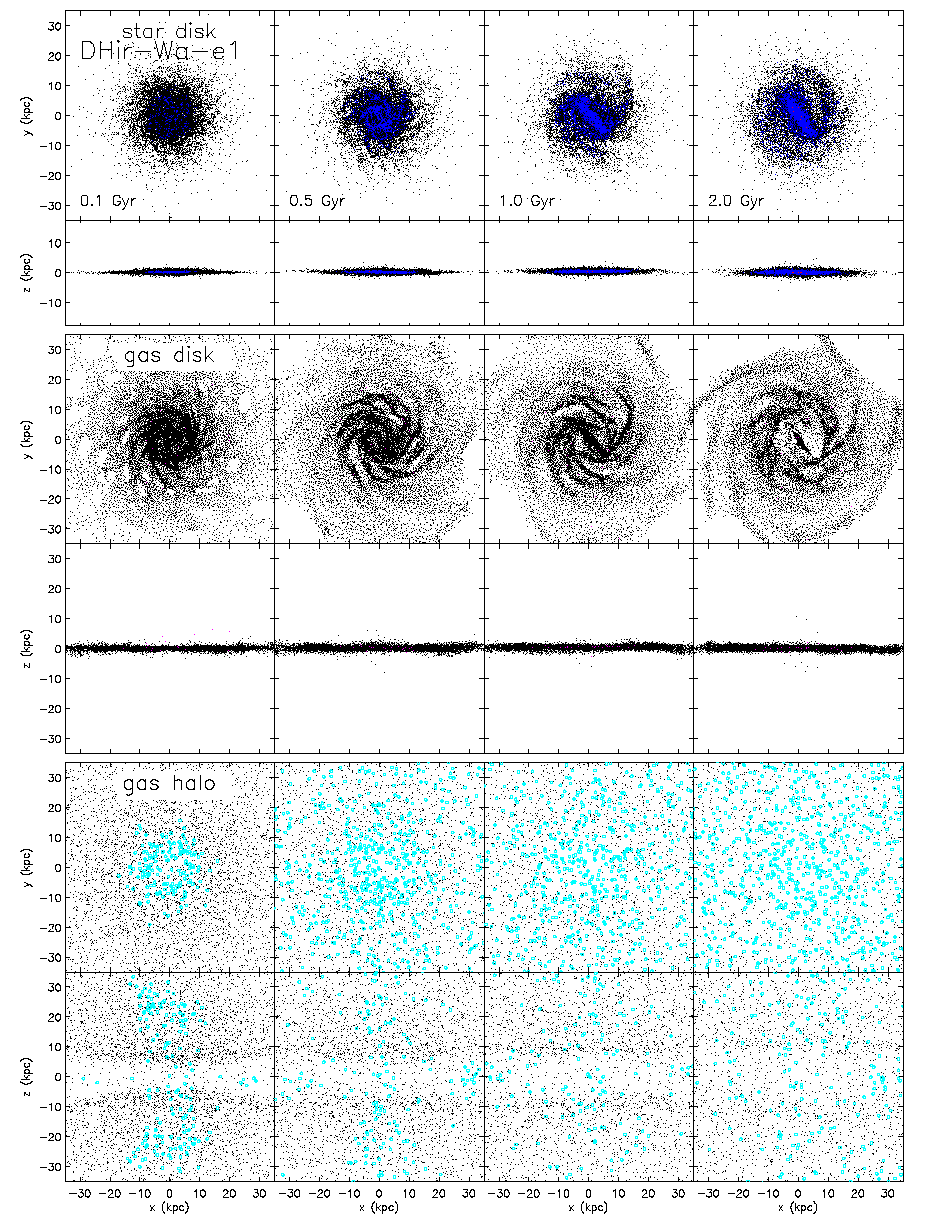}
\caption{
The same snapshots as Fig.~14, but from run~DHir-Wa-el.
As the previous run~DHir-Wa, this run includes axial galactic winds
but with a half the wind efficiency.
}
\end{figure*}

\begin{figure*}[!hbt]
\centering%
\includegraphics[width=14cm]{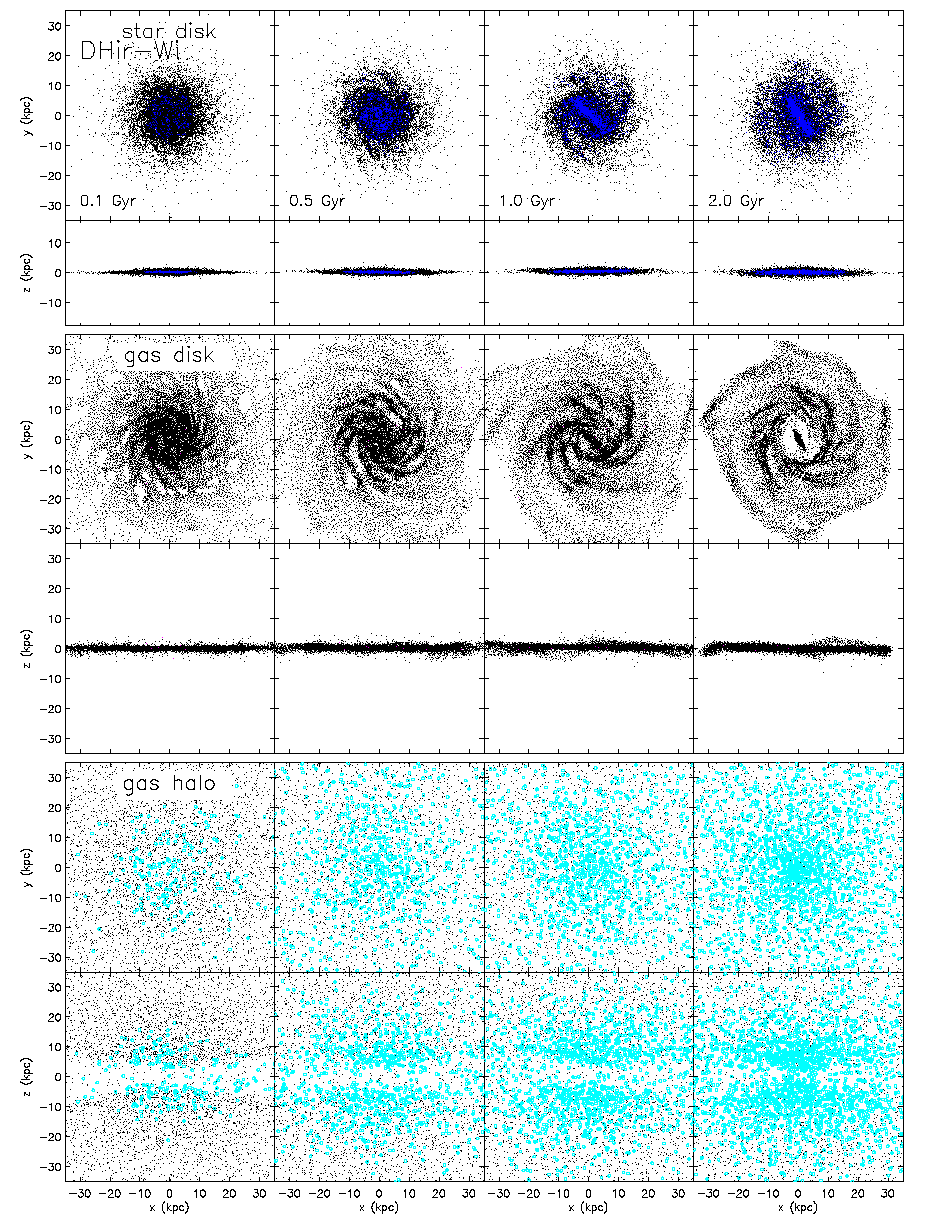}
\caption{
The same snapshots as Fig.~14, but from run DHir-Wi
which includes isotropic galactic winds.
}
\end{figure*}

\begin{figure*}[!hbt]
\centering%
\includegraphics[width=14cm]{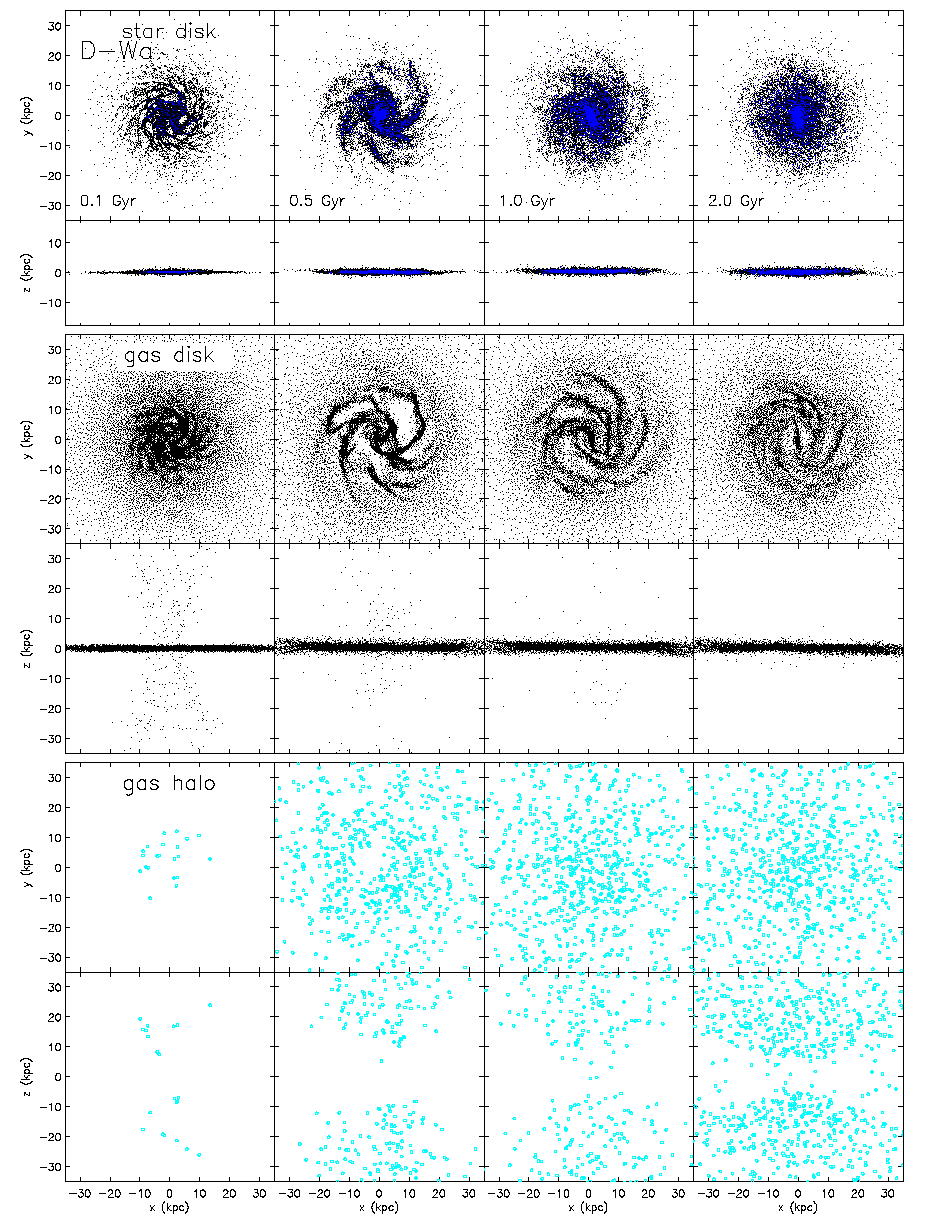}
\caption{
The same snapshots as Fig.~14, but from run~D-Wa
which includes axial galactic winds.
}
\end{figure*}


\subsubsection{Runs DHir-Wa, DHir-Wa-e1, and DHir-Wi}

The galaxy model~DHir is used for these three runs.
The initial model contains a gas halo
with the isothermal density profile and the halo gas fraction of
$f_{\rm hg} = 0.01$ as in the fiducial model~DHi but has an additional rotation
for the gas halo.
The three runs are thus identical to one another
at the start of the runs,
and also to run~DHir without winds in Sec.~3.2.1.
However, they are subject to different evolution
due to the different wind parameters.

The snapshots of run~DHir-Wa at $t$ = 0.1, 0.5, 1, and 2~Gyr
are presented in Fig.~14.
This run accounts for supernova-driven galactic winds in axial mode
with the wind efficiency of 2 as summarized in Table~3.
Compared to run~DHir without winds,
new stars are formed less actively in the star disk
from the gas disk and the gas halo particles
(blue and red dots in the first two rows),
as some gas are blown away by the winds.
Fig.~13 shows that the surface density of the disk
at the center does not rise as high as
that of run~DHir at later times.
The total mass of the star disk grows more slowly
than in run~DHir,
and becomes
$M_{\rm ds}$ = 4.449 $\times$ $10^{10}\,\rm{M_{\odot}}$ at $t$ = 2~Gyr
which is only a 1\% increase from the initial mass (Table~4).
However, the mass of the gas disk is reduced more than in run~DHir.
This is because,
although fewer gas particles in the disk turn into stars,
more gas is expelled from the disk by the energetic winds
(as seen cyan circles in the bottom two rows)
and also fewer gas particles from the halo settle down onto the disk
(magenta dots in the middle two rows).
The phenomena
can be seen in the $T$-$\rho_{\rm local}$ diagrams (Fig.~12)
as well.
In the first row of Fig.~12,
more gas particles originating from the disk (cyan dots)
appear above the criterion line at both $t$~=~1 and 2~Gyr
(first and third panels; see also $N_{\rm hg,d2h}$ in Table~4)
compared to those of run~DHir, and
less gas originating from the halo (magenta dots)
lie below the black line
(second and forth panels; $N_{\rm dg,h2d}$ in Table~4).
The surface density of the gas disk
and the spherically averaged density of the gas halo (Fig.~13)
show less density in the central region at later times
than those of run~DHi,
due to the gas removal from the center by the winds.
The total mass of the gas halo increases
through the gas ejected from the disk
while less amount of gas is lost
to the disk; it becomes
$M_{\rm hg}$ = 1.305 $\times$ $10^{10}\,\rm{M_{\odot}}$ at $t$ = 2~Gyr,
which is a 9\% increase from the initial mass,
while that of the run~DHir
is almost unchanged for a few Gyr.

Four snapshots of run~DHir-Wa-e1
are presented in Fig.~15.
In this simulation, the axial mode of stellar winds
is chosen
as in the previous run~DHir-Wa,
but with half the strength of the wind efficiency
$\eta$ = 1 for comparison.
Mostly, star formation occurs more actively in this run than in run~DHir-Wa;
the total star disk mass increases to
$M_{\rm ds}$ = 4.469 $\times$ $10^{10}\,\rm{M_{\odot}}$ at $t$ = 2~Gyr,
which is a bit greater than that of run~DHir-Wa.
The total mass of the gas disk decreases less slowly than
that in run~DHir-Wa,
because fewer disk gas are expelled from the disk
by the weaker winds
(cyan circles in the snapshots; refer to $N_{\rm hg,d2h}$ in Table~4;
see also the $T$-$\rho_{\rm local}$ plots in Fig.~12),
although more disk gas turns into stars than in run~DHir-Wa
(blue dots; $N_{\rm ds,d2s}$).
The total mass of the gas halo increases to
$M_{\rm hg}$ = 1.271 $\times$ $10^{10}\,\rm{M_{\odot}}$ at $t$ = 2~Gyr,
which is smaller than that of run~DHir-Wa.
The surface density of the gas disk (Fig.~13) rises at the center
at later times
than in run~DHir-Wa.
The spherically averaged density of the gas halo changes
mainly in the inner part.

Four snapshots of run~DHir-Wi are presented in Fig.~16.
This run adopts the isotropic mode for stellar winds.
The wind efficiency is chosen to $\eta$ = 2 as in run~DHir-Wa.
The number of newly formed stars in the disk is
generally similar as that of run~DHir-Wa.
The total masses of the star and the gas disks at 2~Gyr
are also close to those of run~DHir-Wa.
In the snapshots, the expelled gas from the disk
by the winds shows weaker bipolar pattern
(cyan circles; also refer to the $T$-$\rho_{\rm local}$ plots in Fig.~12)
than in run~DHir-Wa,
due to the different wind modes used in the two simulations.
The evolution of surface density profile of the gas disk and
the spherically averaged density profile of the gas halo (Fig.~13)
are not identical to those of run~DHir-Wa, however,
the amount of the ejected gas as well as
the mass of each component are close to
those of run~DHir-Wa (Table~4).


\subsubsection{Run~D-Wa}

This run uses the galaxy model~D,
which does not have a gas halo.
To compare with run~D in Sec.~3.2.3 without winds,
run~D-Wa accounts for galactic winds throughout the simulation.
Axial winds with the wind efficiency $\eta$ = 2
are chosen in this run as in run~DHir-Wa (Table~3).

The snapshots of run~D-Wa at $t$ = 0.1, 0.5, 1, and 2~Gyr
are presented in Fig.~17.
Compared to run~D without winds,
fewer number of stars are added onto the star disk
from the gas disk
(blue dots; also refer to $N_{\rm ds, d2s}$ in Table~4).
The gas flow caused by the winds is seen
as a bipolar pattern in the snapshots.
Some of the gas particles ejected from the plane of the gas disk
are classified as the gas disk particles in the snapshots,
and the others as the gas halo particles (cyan circles),
according to the $T$-$\rho_{\rm local}$ criterion.
At later times, most of the ejected particles
are defined as the halo gas.
The $T$-$\rho_{\rm local}$ diagrams (bottom row in Fig.~12) show that
some of the gas particles originally set as the gas disk
are located above the criterion line due to the wind effects;
this is in contrast to the case of run~D
where no gas particles cross the black line.
As in the other runs with winds,
the winds prevent gas particles from gathering at the center
resulting in the smaller surface density at the center of the disk
compared to run~D,
as seen in Fig.~13.
The total mass of the star disk increases from the initial value
$M_{\rm ds}$ = 4.399 $\times$ $10^{10}\,\rm{M_{\odot}}$
to $M_{\rm ds}$ = 4.455 $\times$ $10^{10}\,\rm{M_{\odot}}$
at $t$ = 2~Gyr (Table~4),
and that of the gas disk decreases from the initial value
$M_{\rm dg}$ = 0.6 $\times$ $10^{10}\,\rm{M_{\odot}}$
to $M_{\rm ds}$ = 0.4824 $\times$ $10^{10}\,\rm{M_{\odot}}$
at $t$ = 2~Gyr.
The number of the gas halo particles
at $t$ = 2~Gyr is $N_{\rm hg}$ = 2071.


\subsection{Star Formation History}

Fig.~18 shows the SFRs of all of our simulations for 3~Gyr.
The left column is for the runs without winds,
and the right column is for the wind test runs.

\begin{figure*}[!hbt]
\centering%
\includegraphics[width=14cm]{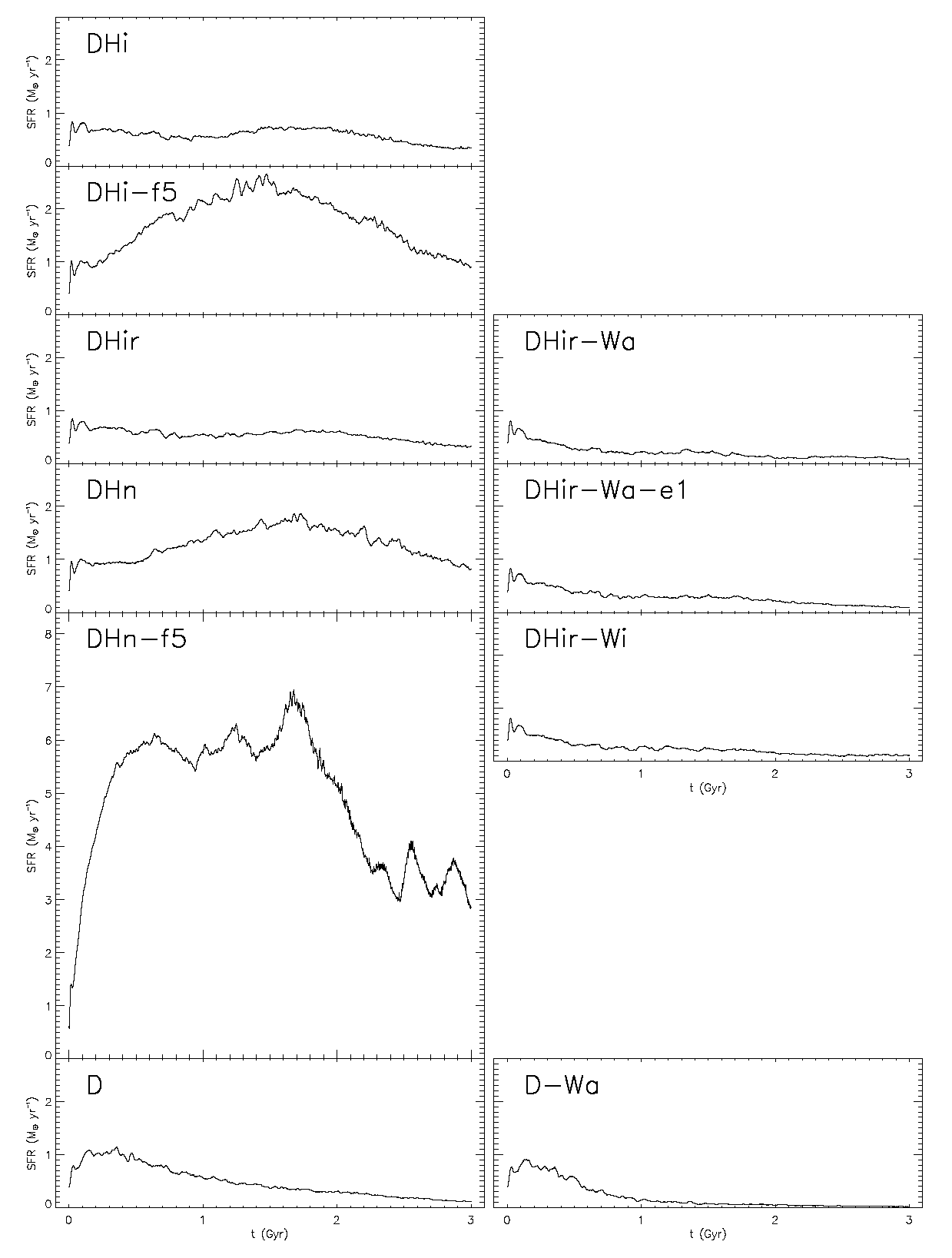}
\caption{
SFRs over the course of simulations.
The left and right columns show the rate from the runs without winds and with winds, respectively.
Run~Hi is not shown as it has zero star formation
throughout the simulation.}
\end{figure*}

In run~D,
where the galaxy model starts with a single gas component of the gas disk
and does not account for the effects of galactic winds,
the SFR (bottom-left panel) increases at the beginning
for about half~Gyr and decreases continuously afterwards.
The maximum rate is 1.14 $\rm{M_{\odot}}~yr^{-1}$ at 0.36~Gyr.
The star formation decreases afterwards
because the cold and dense gas becomes
less available as they are continually consumed by star formation
(refer to Table~4).
In run~Hi, where the initial galaxy model has only a hot gas halo,
the SFR is zero during the entire run.

The SFRs in the runs of type~DH models without winds
(top to fifth panels in the left column),
where the galaxy models are initialized with both gas disk and gas halo,
come out quite differently from that of run~D without a gas halo.
In the fiducial run~DHi
(with an isothermal gas halo of $f_{\rm gh}$=0.01),
the SFR does not change much throughout the simulation.
The rate reaches 0.837 $\rm{M_{\odot}}~yr^{-1}$ at 0.024~Gyr,
decreases slowly to 0.468 $\rm{M_{\odot}}~yr^{-1}$ at 0.91~Gyr,
and increases again to 0.745 $\rm{M_{\odot}}~yr^{-1}$ at 1.49~Gyr.
The SFR is kept above 0.65 $\rm{M_{\odot}}~yr^{-1}$ until about 2~Gyr
and decreases afterwards.
Compared with run~D,
the SFR in the fiducial run
is lower for about the first 1~Gyr.
However, the SFR becomes higher during the last 2~Gyr than in run~D.
Most of the new stars in this run are generated from the gas
originating from the gas disk (Table~4).

In run~DHi-f5 (with a more massive isothermal gas halo of $f_{\rm gh}$=0.05),
the SFR is higher than in the run with the fiducial model~DHi at all times.
The SFR in run~DHi-f5 increases steeply
to the maximum rate 2.658 $\rm{M_{\odot}}~yr^{-1}$ at 1.48~Gyr and
decreases later to 0.9 $\rm{M_{\odot}}~yr^{-1}$ at the end of the run.
The maximum rate is about three times greater than that in the fiducial run.
The time when the rate reaches its maximum is affected
by the adopted value of the model parameter for star formation time-scale
$t_{0}^{\star}$ = 1.47~Gyr.
Star formation occurs mostly in the gas originally set as the gas disk
and only a few percent of the stars are formed
out of the initially halo gas (Table~4).

The SFR in run~DHir (with a rotating isothermal gas halo of $f_{\rm gh}$=0.01)
is maintained relatively constant throughout the simulation
than in all the other runs.
The SFR is very close to that in the fiducial run~DHi
(with the same but non-rotating gas halo)
for the first 1~Gyr but gets lower in the middle of the run
from $\sim$1~Gyr to $\sim$2.5~Gyr by upto $\sim$0.2 $\rm{M_{\odot}}~yr^{-1}$.
As listed in Table~4,
the number of stars generated from the gas initially set as the gas disk
is smaller than that in the fiducial run at each time;
the number of new stars from the gas originally from the gas halo
is a few times larger, however their total mass is still insignificant
as $\lesssim$ 1\% of the total mass of all new stars.
It is implied that the rotation of the gas halo in this run
helps some halo gas particles to be accreted down to the gas disk more easily,
however, it affects overall to lower star formation activity
than in the run without the initial rotation.

In run~DHn (with a NFW gas halo of $f_{\rm gh}$=0.01),
the SFR is much higher than that in the fiducial run~DHi
at all times due to the more centrally concentrated NFW gas halo.
The SFR in run~DHn is rather more comparable to that in run~DHi-f5
than in run~DHi,
although the maximum rate 1.866 $\rm{M_{\odot}}~yr^{-1}$ at 1.72~Gyr
is certainly lower than in run~DHi-f5.
The percentage of the stars formed out of the gas originating from the gas halo
becomes significant in run~DHn;
it is a few times greater than that in run~DHi-f5 (Table~4).

In run~DHn-f5 (with a massive NFW gas halo of $f_{\rm gh}$=0.05),
the SFR is higher than in all the other runs throughout the simulation.
The SFR increases very rapidly from the start of the run
over 5.5 $\rm{M_{\odot}}~yr^{-1}$ in 0.35~Gyr,
and reaches the maximum rate 6.94 at 1.66~Gyr.
After the maximum, it decreases rapidly
to 2.960 $\rm{M_{\odot}}~yr^{-1}$ at 2.47~Gyr
and slows down till the end of the run.
The amount of new stars generated from the halo gas at later times
is comparable to that from the disk gas,
as the gas set as the gas disk are largely consumed by the times
while a significant number of gas particles
from the centrally concentrated massive gas halo
are still cooled down to the gas disk (Table~4).

In the wind test runs,
the SFRs come out always lower
than those in the corresponding runs without winds,
because the winds disturb cold gas from being concentrated.
Including axial winds with the wind efficiency $\eta$ = 2,
the SFR in run~D-Wa (bottom-right) has
its maximum 0.91 $\rm{M_{\odot}}~yr^{-1}$ at 0.13~Gyr,
which is 0.23 $\rm{M_{\odot}}~yr^{-1}$ less than the maximum rate in run~D.
After the maximum, the SFR decreases more rapidly than in run~D
down to 0.147 $\rm{M_{\odot}}~yr^{-1}$ at 1~Gyr and 0.022 at 3~Gyr.
In run~DHir-Wa with axial winds and the efficiency $\eta$ = 2,
the SFR is also lower than in run~DHir without winds.
With the reduced wind efficiency $\eta$ = 1 in run~DHir-Wa-e1,
the SFR becomes overall higher than that in run~DHir-Wa
and lower than that in run~DHir.
In run~DHir-Wi adopting isotropic winds
with the wind efficiency $\eta$ = 2,
the SFR comes out close to that in run~DHir-Wa with axial winds.


\section{SUMMARY AND DISCUSSION}

Motivated by the existence of hot gaseous halo in spiral galaxies
evidenced by both observations and simulations,
we have investigated the effects of a gas halo on the evolution of
isolated Milky Way-like galaxies
using $N$-body/Hydrodynamic simulations.
We have also examined the effects of galactic winds driven by supernovae
on the evolution.

To generate the initial galaxy models,
we used the ZENO software package.
We built seven different galaxy models,
with or without a gas halo component
with varying mass and density profile for comparison (Tables~1 and 2):
We tried two different density profiles for the gas halo -
an isothermal profile
and an NFW profile.
We also tested different amounts of the halo gas fraction -
$f_{\rm hg}$ = 0.01
and $f_{\rm hg}$ = 0.05.
In addition, we studied a case with an initially rotating gas halo.
We set the disk gas fraction $f_{\rm dg}$ = 0.12 equally
in all models possessing a gas disk.

Each galaxy model was then evolved in isolation
using the $N$-body/SPH code GADGET-3 (an early version),
first without including stellar winds driven by star formation.
The effects of winds are also studied.
In the wind test runs, either axial winds
or isotropic winds
were tested, with the wind efficiency of either $\eta$ = 2
or $\eta$ = 1.

We showed that the effects of the hot gas component
on the evolution are significant,
particularly in the gas disk dissipation/accretion and
in the star formation activity.
The runs with a gas halo
show higher SFRs in the middle to last phase
of the 3~Gyr-simulations
than in the run without a gas halo
where the SFR reaches its maximum at the early phase
and then decreases continuously thereafter.
The star formation activity in the disks
is affected by both the mass concentration at the inner part
and the total mass of the gas halo.
The more centrally concentrated NFW gas halo leads
higher SFRs than in runs with the isothermal gas halo
of the same total mass.
In runs
where the gas halo has higher density at the central halo
and/or greater total mass,
the SFRs are higher than others and
the gas disks dissipate too quickly.
The gas accretion from the halo onto the disk in those runs
turns out to occur not much actively,
and some portion of the accreted gas are quickly consumed
by star formation as well,
resulting in unrealistically small gas disks after a few Gyr.
It is not clear which physical processes are mainly responsible
for the dissipation of gas disks.

Models~DHi (the fidicial model)
and DHir (with the additional rotation of the gas halo
to the fiducial model) show relatively reasonable results
in both the gas dissipation and the star formation activity.
The SFR in run~DHi
is a bit higher than that in run~DHir in the middle of the evolution.
The mass of the gas disk is slightly greater in run~DHir.
It turned out that the initial rotation of the gas halo
leads more gas accretion from the halo to the disk
but delays the star formation activity in the disk.
The specific mechanism of how the rotation hinders star formation,
however, needs to be further investigated.

With model~DHir and model~D
we performed four more simulations including supernova-driven galactic winds.
In these wind test runs,
the SFRs come out lower than
the corresponding runs without winds at all times as expected.
Both axial
and isotropic
winds result in overall similar SFRs.
The run with the reduced wind efficiency
shows higher SFR than that with the default value of the wind efficiency.

In our initial galaxy models,
we kept the total mass of the system as well as the total masses of
the halo (DM + gas), disk (stars + gas), and bulge components equally
to focus on examining the impacts of
the different gas distribution and rotation of the gas halo.
Our simulation results, such as in the gas dissipation and accretion and
the star formation activities, are restricted
by our initial galaxy models and
the parameter values in the simulation code.
The wind effects would also be affected
by a variety of factors, for instance the total mass of the galaxy model,
and be limited by the provided wind model in the code.

Nonetheless,
numerical studies to seek more realistic galaxy models
including a gas halo component are demanded
since the component as a large reservoir of gas influences
the evolution of the galaxies.
Adding a gas halo in numerical simulations,
various gas dynamical processes that may occur in real systems
both isolated or interacting with others,
would be better interpreted.

The galaxy models are eventually compared with
the `galaxies' in cosmological simulations \citep{Kim2011}.
We plan to use our isolated galaxy models to understand
the morphology and luminosity transformation of interacting galaxies \citep{Park2008}
found in the Sloan Digital Sky Survey \citep{Choi2010}.


\begin{acknowledgments}
We are grateful to
Joshua E. Barnes and Volker Springel for offering
the ZENO package and the GADGET code, respectively.
J.-S. H. thanks Walter Dehnen, Juhan Kim, and
Maurice H. P. M. van Putten for helpful discussions.
We thank the Korea Institute for Advanced Study
for providing computing resources (KIAS
Center for Advanced Computation Linux Cluster System)
for this work.
\end{acknowledgments}

\begin{figure*}[!hbt]
\centering%
\includegraphics[width=14cm]{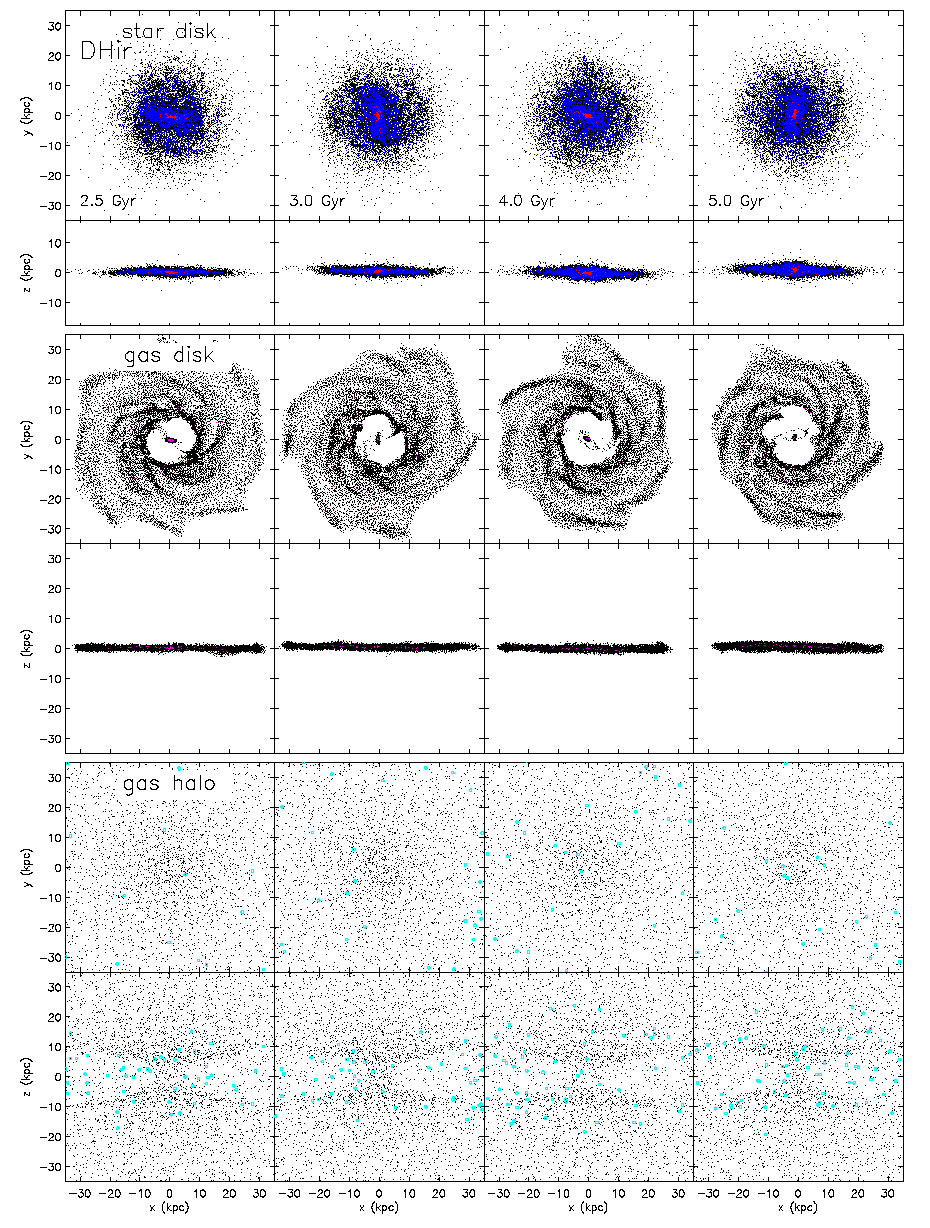}
\caption{
The particle distribution of run~DHir without winds after the time shown in Fig.~7.
The first to fourth columns show the snapshots at $t$~=~2.5, 3, 4, and 5~Gyr, respectively.
}
\end{figure*}
%


\appendix
\subsection{A. Construction of initial galaxy models with ZENO}

Here we describe the procedure for constructing
the models possessing both gas disk and gas halo
and then disk-only or halo-only models,
using several programs provided in the ZENO package (c.f. \citealt{Barnes2009}).

To construct the $N$-body realization of the spherical components
of the models of type~DH,
we first obtain the cumulative mass profiles of each component
by integrating the corresponding mass density profiles
with the parameter values shown in Table~4.

For the bulge, we use the program {\it gammagsp} in the ZENO package.
This code generates profiles for Dehnen models \citep{Dehnen1993},
including the Hernquist profile (Eq.~4)
with $gamma=1$ in the input options.
We truncate the mass distribution of the bulge
from the radius $b_{\rm b}$ as indicated in Eq.~4
and rescale to the original mass, using the program {\it gsptrun}.
For the profile of either DM or gas halo
following the NFW model (Eq.~2), we run {\it halogsp}.
The profile of the gas halo following the non-singular isothermal
form (Eq.~3; type~DHi) is generated by {\it isothgsp}.
The total spheroid mass profile and the total mass profile
are calculated in accordance with the choice of the halo profile.
The disk is represented by an infinitely thin disk
at this stage (see \citealt{Barnes2009} for detail),
and the profile of the `spherical' (infinitely thin)
disk is generated by {\it expdgsp}.
We then combine the mass profiles of the bulge, the DM and the gas halos,
and the `spherical' disk to get
the total cumulative mass profile using {\it gspadd},
and smooth the profile using {\it gspsmooth}.
Finally, we generate each of the spherical components,
running two $N$-body realization codes,
{\it gsprealize} for the bulge and the DM halo
and {\it gspsphere} for the gas halo,
with the input parameters of the density profile of the component as well as
the smoothed total mass profile for the gravity calculation.
The {\it gsprealize} code
uses the Abel integral to compute the distribution function and
generates a configuration of the body in phase-space.
The {\it gspshere} code outputs
the initial positions and internal energies of the halo gas particles
with zero initial velocities.

To make realizations of the star and the gas disks
we run {\it gspdisk} twice. This code makes an exponential disk
embedded in the smoothed total spheroid mass profile
with the disk mass and other parameters given in Table~4.
For the initial velocities of the star disk particles,
the local circular velocities and the vertical and radial dispersions
are calculated in the way described in \citealt{Barnes2009}.
For the gas particles, we set zero dispersion.

The procedure for making model~DHir (disk plus halo model with rotation)
is the same as in model~DHi (the fiducial model),
except an additional step for giving rotation
of the gas halo using a simple code we made.
We set the rotation such that the gas halo has the rotation curve
with a similar shape but half the amplitude of that of the disk.

To build either the disk-only model (model~D)
or the halo-only model (model~Hi),
one can follow the steps described above
but skipping the parts related to
either the gas halo or the gas disk, respectively.

\subsection{B. Long-term evolution of model DHir}

Fig.~19 shows
four more snapshots from run~DHir at later times since those in Fig.~7.
More stars are seen in the star disk than in Fig.~7.
The inner part of the gas disk appears largely dissipated
as time passes,
and the refilling of the gas disk by accreting halo gas
does not seem to occur actively.

The total mass of the star disk increases to
$M_{\rm ds}$ = 4.56 $\times$ $10^{10}\,\rm{M_{\odot}}$ at $t$ = 3~Gyr,
and 4.59 $\times$ $10^{10}\,\rm{M_{\odot}}$ at $t$ = 5~Gyr.
Among the stars in the disk at $t$ = 3~Gyr,
the number of stars added from the gas
originally set as the gas disk and the gas halo are
$N_{\rm ds,d2s}$ = 8600 (blue dots in the first two rows)
and $N_{\rm ds,h2s}$ = 94 (red dots), respectively.
The total mass of the gas disk decreases to
$M_{\rm dg}$ = 0.44 $\times$ $10^{10}\,\rm{M_{\odot}}$ at $t$ = 3~Gyr,
and 0.41 $\times$ $10^{10}\,\rm{M_{\odot}}$ at $t$ = 5~Gyr.
The total mass of the gas halo is maintained to
$M_{\rm hg}$ = 1.2 $\times$ $10^{10}\,\rm{M_{\odot}}$ at $t$ = 3 and 5~Gyr.



\begin{thebibliography}{}

\bibitem[Anderson \& Bregman(2010)]{Anderson2010}
		 Anderson, M. E., \& Bregman, J. N.\ 2010,
		 Do Hot Halos Around Galaxies Contain the Missing Baryons?,
		 \apj, 714, 320
\bibitem[Barnes \& Hernquist(1992)]{Barnes1992}
         Barnes, J.~E., \& Hernquist, L.\ 1992,
         Dynamics of Interacting Galaxies,
         \araa, 30, 705
\bibitem[Barnes \& Hibbard(2009)]{Barnes2009}
		 Barnes, J. E., \& Hibbard, J. E.\ 2009,
		 Identikit 1: A Modeling Tool for Interacting Disk Galaxies,
		 \aj, 137, 3071
\bibitem[Blumenthal et al.(1984)]{Blumenthal1984}
         Blumenthal, G.~R., Faber, S.~M., Primack, J.~R., \& Rees, M.~J.\ 1984,
         Formation of Galaxies and Large-Scale Structure with Cold Dark Matter,
         \nat, 311, 517
\bibitem[Choi \& Nagamine(2011)]{Choi2011}
         Choi, J.-H., \& Nagamine, K.\ 2011,
         Multicomponent and Variable Velocity Galactic Outflow in
         Cosmological Hydrodynamic Simulations,
         \mnras, 410, 2579
\bibitem[Choi et al.(2010)]{Choi2010}
	     Choi, Y.-Y., Han, D.-H., \& Kim, S. S.\ 2010,
         Korea Institute for Advanced Study Value-Added Galaxy Catalog,
         JKAS, 43, 191
\bibitem[Dehnen(1993)]{Dehnen1993}
         Dehnen, W.\ 1993,
	     A Family of Potential-Density Pairs for Spherical Galaxies and Bulges,
	     \mnras, 265, 250
\bibitem[Dekel \& Silk(1986)]{Dekel1986}
         Dekel, A., \& Silk, J.\ 1986,
         The Origin of Dwarf Galaxies, Cold Dark Matter, and Biased Galaxy Formation,
         \apj, 303, 39
\bibitem[Deng et al.(2012)]{Deng2012}
         Deng, X.-F., Zhang, F., Song, J., Chen, Y.-Q., \& Jiang, P. \ 2012,
         The Environmental Dependence of the Fraction of `Unconventional' Galaxies: Faint Red and Luminous Blue,
         JKAS, 45, 59
\bibitem[Hernquist(1990)]{Hernquist1990}
		 Hernquist, L.\ 1990,
		 An Analytical Model for Spherical Galaxies and Bulges,
		 \apj, 356, 359
\bibitem[Hernquist(1993)]{Hernquist1993}
         Hernquist, L.\ 1993,
         $N$-body Realizations of Compound Galaxies,
         \apjs, 86, 389
\bibitem[Hopkins et al.(2011)]{Hopkins2011}
         Hopkins, P.~F., Quataert, E., \& Murray, N.\ 2011,
         Self-Regulated Star Formation in Galaxies via Momentum Input from Massive Stars,
         \mnras, 417, 950
\bibitem[Hwang \& Park(2009)]{Hwang2009}
         Hwang, H.~S., \& Park, C. \ 2009,
         Evidence for Morphology and Luminosity Transformation of Galaxies at High Redshifts,
         \apj, 700, 791
\bibitem[Katz et al.(1996)]{Katz1996}
		 Katz, N., Weinberg, D. H., \& Hernquist, L.\ 1996,
		 Cosmological Simulations with TreeSPH,
		 \apjs, 105, 19
\bibitem[Kazantzidis et al.(2004)]{Kazabtzidis2004}
         Kazantzidis, S., Magorrian, J., \& Moore, B.\ 2004,
         Generating Equilibrium Dark Matter Halos: Inadequacies of
         the Local Maxwellian Approximation,
         \apj, 601, 37
\bibitem[Kennicutt(1998)]{Kennicutt1998}
    	 Kennicutt, R. C. Jr.\ 1998,
		 The Global Schmidt Law in Star-Forming Galaxies,
		 \apj, 498, 541
\bibitem[Kim et al.(2011)]{Kim2011}
         Kim, J., Park, C., Rossi, G., Lee, S. M., \& Gott, J. R. \ 2011,
         The New Horizon Run Cosmological $N$-Body Simulations,
         JKAS, 44, 217 		
\bibitem[McKee \& Ostriker(1977)]{Mckee1977}
         McKee, C. F., \& Ostriker, J. P.\ 1977,
         A Theory of the Interstellar Medium: Three Components
         Regulated by Supernova Explosions in an Inhomogeneous Substrate,
         \apj, 218, 148
\bibitem[McMillan \& Dehnen(2007)]{McMillan2007}
		 McMillan, P. J., \& Dehnen, W.\ 2007,
		 Initial Conditions for Disc Galaxies,
		 \mnras, 378, 541
\bibitem[Merritt(1985)]{Merritt1985}
         Merritt, D.\ 1985,
         Distribution Functions for Spherical Galaxies,
         \mnras, 214, 25P
\bibitem[Moster et al.(2011)]{Moster2011}
		 Moster, B. P., Macci\`o, A. V., Somerville, R. S., Naab, T. \& Cox, T. J.\ 2011,
		 The Effects of a Hot Gaseous Halo in Galaxy Major Mergers,
		 \mnras, 415, 3750
\bibitem[Moster et al.(2012)]{Moster2012}
		 Moster, B. P., Macci\`o, A. V., Somerville, R. S., Naab, T. \& Cox, T. J.\ 2012,
		 The Effects of a Hot Gaseous Halo on Disc Thickening in Galaxy Minor Mergers,
		 \mnras, 423, 2045
\bibitem[Navarro et al.(1996)]{Navarro1996}
		 Navarro, J. F., Frenk, C. S., \& White, S. D. M.\ 1996,
		 The Structure of Cold Dark Matter Halos,
   	 	 \apj, 462, 563
\bibitem[Osipkov(1979)]{Osipkov1979}
         Osipkov, L.~P.\ 1979,
         Spherical Systems of Gravitating Bodies with
         an Ellipsoidal Velocity Distribution,
         Pis'ma Astr. Zh., 5, 77
\bibitem[Park \& Choi(2009)]{Park_Choi2009}
         Park, C., \& Choi, Y.-Y. \ 2009,
         Combined Effects of Galaxy Interactions and Large-Scale Environment on Galaxy Properties,
         \apj, 691, 1828
\bibitem[Park et al.(2008)]{Park2008}	
	     Park, C., Gott, J. R., \& Choi, Y.-Y.\ 2008,
	     Transformation of Morphology and Luminosity Classes of the SDSS Galaxies,
	     \apj, 674, 784
\bibitem[Park \& Hwang(2009)]{Park_Hwang2009}
         Park, C., \& Hwang, H.~S. \ 2009,
         Interactions of Galaxies in the Galaxy Cluster Environment,
         \apj, 699, 1595
\bibitem[Salpeter(1955)]{Salpeter1955}
		 Salpeter, E. E.\ 1955,
		 The Luminosity Function and Stellar Evolution,
    	 \apj, 121, 161
\bibitem[Springel(1990)]{Springel1990}
		 Springel, V.\ 2005,
		 The Cosmological Simulation Code GADGET-2,
		 \mnras, 364, 1105
\bibitem[Springel et al.(2005)]{Springel2005}
         Springel, V., Di Matteo, T., \& Hernquist, L.\ 2005,
         Modelling Feedback from Stars and Black Holes in Galaxy Mergers,
         \mnras, 361, 776
\bibitem[Springel \& Hernquist(2002)]{Springel2002}
		 Springel, V., \& Hernquist, L.\ 2002,
		 Cosmological Smoothed Particle Hydrodynamics Simulations: The Entropy Equation,
		 \mnras, 333, 649
\bibitem[Springel \& Hernquist(2003)]{Springel2003}
		 Springel, V., \& Hernquist, L.\ 2003,
		 Cosmological Smoothed Particle Hydrodynamics Simulations:
		 A Hybrid Multiphase Model for Star Formation,
		 \mnras, 339, 289
\bibitem[Toomre \& Toomre(1972)]{Toomre1972}
         Toomre, A., \& Toomre, J.\ 1972,
         Galactic Bridges and Tails,
         \apj, 178, 623
\end{thebibliography}
\end{document}